\documentclass[12pt, preprint]{aastex}
\usepackage{color}

\usepackage[normalem]{ulem}
\usepackage{natbib}
\usepackage{hyperref}
\usepackage{cleveref}

\def\esym{$E_{\rm{sym}}(\rho)$~}
\shorttitle{Combined Constraints on the Equation of State of Dense Neutron-Rich Matter from Terrestrial Experiments and Observations of Neutron Stars}
\shortauthors{Zhang et al.}
\begin{document}
\title{Combined Constraints on the Equation of State of Dense Neutron-Rich Matter from Terrestrial Nuclear Experiments and Observations of Neutron Stars}
\author{Nai-Bo Zhang\altaffilmark{1,2}, Bao-An Li\altaffilmark{1,*}, and Jun Xu\altaffilmark{3}}

\altaffiltext{1}{Department of Physics and Astronomy, Texas A$\&$M University-Commerce, Commerce, TX 75429, USA}
\altaffiltext{2}{Shandong Provincial Key Laboratory of Optical Astronomy and Solar-Terrestrial Environment,
Institute of Space Sciences, Shandong University, Weihai, 264209, China}
\altaffiltext{3}{Shanghai Institute of Applied Physics, Chinese Academy of Sciences, Shanghai 201800, China}
\altaffiltext{*}{Corresponding author: Bao-An.Li@Tamuc.edu}

\begin{abstract}
Within the parameter space of equation of state (EOS) of dense neutron-rich matter limited by existing constraints mainly from terrestrial nuclear experiments, we investigate how the neutron star maximum mass
$M_{\rm{max}}>2.01\pm0.04$ M$_\odot$, radius $10.62<R_{\rm{1.4}}< 12.83$ km and tidal deformability $\Lambda_{1.4}\leq800$ of canonical neutron stars all together constrain the EOS of dense neutron-rich nucleonic matter.
While the 3-D parameter space of $K_{\rm{sym}}$ (curvature of nuclear symmetry energy), $J_{\rm{sym}}$ and $J_0$ (skewness of the symmetry energy and EOS of symmetric nuclear matter, respectively)
are narrowed down significantly by the observational constraints, more data are needed to pin down the individual values of $K_{\rm{sym}}$, $J_{\rm{sym}}$ and $J_0$ with quantified uncertainties. The $J_0$ largely controls the maximum mass of neutron stars. While the EOS with $J_0=0$ is sufficiently stiff to support neutron stars as massive as 2.37 M$_{\odot}$, to support the hyperthetical ones as massive as 2.74 M$_{\odot}$ (composite mass of GW170817) requires $J_0$ to be larger than its currently known maximum value of about 400 MeV and beyond the causality limit. The upper limit on the tidal deformability of $\Lambda_{1.4}=800$ from the recent observation of GW170817 is found to provide upper limits on some EOS parameters consistent with but far less restrictive than the existing constraints of other observables studied.
\end{abstract}
\maketitle
\newpage
\section{Introduction}
\label{sec1}
What is the nature of neutron stars and dense nuclear matter? To answer this question has been a longstanding and shared goal of both astrophysics and nuclear physics. The fundamental importance and broad impacts of answering this question have been well documented in the literature. It is a major scientific thrust for many major research facilities in astrophysics \citep{NAP2011} and nuclear physics \citep{NAP2012}, such as, various advanced X-ray satellites and earth-based large telescopes, the Neutron Star Interior Composition Explorer (NICER), various gravitational wave detectors and all advanced radioactive beam facilities being built around the world. In particular, to answer this question has been identified as a major goal in both the U.S. 2015 Long Range Plan for Nuclear Sciences \citep{LRP2015} and the Nuclear Physics European Collaboration Committee (NuPECC) 2017 Long Range Plan \citep{NuPECC}. However, answering this question accurately is very challenging. It is well known that properties of neutron stars are determined by the Equation of State (EOS) of neutron-rich matter over a large density range from zero to about ten times normal nuclear matter density. Unfortunately,  even within the simplest model considering $npe\mu$ particles only, the corresponding EOS of neutron star matter is still poorly known, not to mention possibly other particles and various phase transitions predicted to occur in the core of neutron stars.

To summarize what terrestrial experiments and nuclear theories have taught us about the EOS of neutron-rich matter near the saturation density $\rho_0$ of symmetric nuclear matter (SNM), we first recall that 
the specific energy $E(\rho,\delta)$ in asymmetric nucleonic matter (ANM) can be approximated parabolically in isospin asymmetry $\delta=(\rho_{\rm{n}}-\rho_{\rm{p}})/\rho$ as
\begin{equation}\label{PAEb}
  E(\rho,\delta)\approx E_0(\rho)+E_{\rm{sym}}(\rho)\delta^2
\end{equation}
where $E_0(\rho)$ is the specific energy in SNM and $E_{\rm{sym}}(\rho)$ is the symmetry energy.  Around $\rho_0$, the $E_0(\rho)$ and $E_{\rm{sym}}(\rho)$ can be Taylor expanded up to $[(\rho-\rho_0)/3\rho_0]^3$ as
\begin{eqnarray}\label{E0para}
  E_{0}(\rho)&\approx&E_0(\rho_0)+\frac{K_0}{2}(\frac{\rho-\rho_0}{3\rho_0})^2+\frac{J_0}{6}(\frac{\rho-\rho_0}{3\rho_0})^3,\\
  E_{\rm{sym}}(\rho)&\approx&E_{\rm{sym}}(\rho_0)+L(\frac{\rho-\rho_0}{3\rho_0})+\frac{K_{\rm{sym}}}{2}(\frac{\rho-\rho_0}{3\rho_0})^2+\frac{J_{\rm{sym}}}{6}(\frac{\rho-\rho_0}{3\rho_0})^3\label{Esympara}
\end{eqnarray}
in terms of several EOS characteristic parameters: the incompressibility $K_0$ and skewness  $J_0$ of SNM as well as the slope $L$, curvature $K_{\rm{sym}}$ and skewness $J_{\rm{sym}}$ of the symmetry energy.
While the  above Taylor expansions are not used in our study for neutron stars, they provide asymptotic boundary conditions at $\rho_0$ for the parameterizations we shall use to describe the EOS of neutron star matter.

Generally speaking, the EOS of neutron-rich nucleonic matter remains very uncertain mostly because of the poorly known nuclear symmetry energy especially at supra-saturation densities \citep{Tesym}.
Nevertheless, thanks to the hard work of many people in both astrophysics and nuclear physics over many years, much progress has been made in constraining the EOS of neutron-rich nucleonic matter. In particular,
various analyses of terrestrial nuclear experiments and astrophysical observations have constrained the $K_0$, $E_{\rm sym}(\rho_0)$ and $L$ in reasonably small ranges around $K_0\approx 240 \pm 20$ MeV \citep{Shlomo06,Piekarewicz10}, $E_{\rm sym}(\rho_0)=31.7\pm 3.2$ MeV and $L\approx 58.7\pm 28.1 $ MeV. Especially worth noting, the quoted most probable values of $E_{\rm sym}(\rho_0)$ and $L$ are based on surveys of 53 analyses of different kinds of terrestrial and astrophysical data available up to Oct. 2016 \citep{Li13,Oertel17}. Moreover, extensive analyses of heavy-ion reactions at intermediate and relativistic energies, especially various forms of nucleon collective flow and kaon production, have provided a reasonably tight constraining band for the EOS of SNM up to about $4.5\rho_0$, see, e.g., ref.  \cite{Danielewicz02}. However, the coefficients characterizing the high-density behavior of neutron-rich matter, such as the  $K_{\rm{sym}}$, $J_{\rm{sym}}$ and $J_0$ are only loosely known to be in the range of $-400 \leq K_{\rm{sym}} \leq 100$ MeV, $-200 \leq J_{\rm{sym}}\leq 800$ MeV, and $-800 \leq J_{0}\leq 400$ MeV mostly based on analyses of terrestrial nuclear experiments and energy density functionals \citep{Tews17,Zhang17}, respectively. 

While continuous efforts have been made to constrain the EOS of dense neutron-rich matter using heavy-ion reactions which may involve rare isotopes with large neutron/proton ratios, presently no concensus has been reached yet from analyzing limited data available \citep{Li17}. On the other hand, properties of neutron stars and events involving them, such as the mass, radius, moment of inertia, quadrupole deformation, pulsing frequency \citep{Hessels06}, cooling curve \citep{Yakovlev01,Page06}, frequencies and damping times of various oscillating modes, spin parameter of pulsars, as well as the strain amplitude and phase evolution of gravitational waves from inspiraling neutron star binaries all depend significantly on the EOS of neutron-rich matter, see, e.g., refs. \citep{Lattimer16,Oertel17,Ozel16,Watts16,Gri16,Newton14} for recent reviews. While the observational data of neutron star properties are relatively limited so far, they also provide stringent constraints on the EOS and guide theories of dense neutron-rich matter. In particular, the observed masses around 2M$_\odot$ for  the two most massive pulsars J1614-2230 \citep{Demorest10} and J0348+0432 \citep{Antoniadis13} restrict mostly the stiffness of dense SNM EOS, while the radii of neutron stars are known to be more sensitive to the symmetry energy around $2\rho_0$ \citep{Lattimer00,Lattimer01}. While much progress has been made in measuring the radii of neutron stars, because of the great difficulties involved especially in determining accurately the distance and modeling reliably the spectrum absorptions with different atmosphere models, the reported radii normally suffer from relatively large uncertainties. In fact, some radii extracted from different analyses and observations are still controversial, see, e.g., ref. \citep{CMiller} for a recent review. Since we are not in a position to make any judgement on the reliability of any astrophysical observations, in our analysis here we use as an example the radius of $1.4$ M$_\odot$ canonical neutron stars ($R_{\rm{1.4}}$) in the range of $10.62<R_{\rm{1.4}}< 12.83$ km inferred from analyzing quiescent low-mass X-ray binaries in ref. \citep{Lattimer14}. Moreover, we also use the upper limit of the dimensionless tidal deformation $\Lambda_{1.4}\leq800$ from the recent observation of GW170817 by the LIGO+Virgo Collaborations \citep{Abbott17,Ligo17}.

The first discovery of a neutron star merger using multiple messengers further signifies and sets an excellent example of combining and cross-checking multiple probes of dense neutron-rich matter on earth and in heaven.
Given the aforementioned constraints on the EOS parameters of neutron-rich matter mostly based on terrestrial nuclear laboratory experiments, here we study how the astrophysical observations of
$M_{\rm{max}}>2.01\pm0.04$ M$_\odot$, $10.62<R_{\rm{1.4}}< 12.83$ km and $\Lambda_{1.4}\leq800$ all together constrain the high-density EOS in a way consistent naturally with the existing constraints from terrestrial nuclear experiments. For this purpose, fixing the $K_0$, $E_{\rm sym}(\rho_0)$ and $L$ at their most probably values mentioned earlier, we explore the intersections of constant surfaces with $M_{\rm{max}}=2.01$ $\rm{M}_\odot$, $R_{\rm{1.4}}=10.62$ km, $R_{\rm{1.4}}=12.83$ km, and $\Lambda_{1.4}=800$, respectively, in the 3-dimensional (3-D) parameter space of $K_{\rm{sym}}$, $J_{\rm{sym}}$ and $J_0$. The 3-D parameter space allowed by all three observational constraints are identified. Moreover, in constructing the EOS of neutron star matter, the crust-core transition density $\rho_t$ and pressure $P_t$ have to be calculated consistently. While effects of the magnitude $E_{\rm{sym}}(\rho_0)$ and slope $L$ of symmetry energy at $\rho_0$ on  the crust-core transition have been extensively studied in the literature, effects of the curvature $K_{\rm{sym}}$ are less known. We therefore will also explore contours of the $\rho_t$ and $P_t$ in the $K_{\rm{sym}}$ versus $L$ plane (2-D). The significant role of the $K_{\rm{sym}}$ is clearly revealed. Furthermore, within the currently known uncertainty ranges of $J_0$ and $J_{\rm{sym}}$, by setting $J_0=J_{\rm{sym}}=0$ in Eqs. (\ref{E0para}) and (\ref{Esympara}) as often done in the literature, we also explore how/what the same three astrophysical constraints may teach us about the EOS of dense neutron-rich matter in the $L-K_{\rm{sym}}$ parameter plane. In both the 3-D and 2-D model frameworks, the upper limit of the tidal deformability $\Lambda_{1.4}=800$ from the recent observation of GW170817 is found to provide upper limits on some EOS parameters consistent with but less restrictive than the existing constraints on them. Overall, while combing exiting constraints from both terrestrial nuclear experiments and astrophysical observations allows us to limit significantly the EOS parameter space of high-density neutron-rich matter, data of more independent observables are needed to pin down the individual values of $K_{\rm{sym}}$, $J_{\rm{sym}}$ and $J_0$ with quantified uncertainties.

This paper is organized as follows. The construction of the EOS of neutron star matter is presented in Section \ref{sec2}. The Section \ref{sec3} is devoted to constraining the EOS of dense neutron-rich nucleonic matter with astrophysical observations first in the 3-D space of $K_{\rm{sym}}$, $J_{\rm{sym}}$ and $J_0$, then in the 2-D plane of $L-K_{\rm{sym}}$ with $J_{\rm{sym}}=J_0=0$.
Finally, a summary is given in Section \ref{sec4}.

\section{Modeling the equation of state of neutron star matter}
\label{sec2}
The focus of this study is on the EOS of dense neutron-rich matter. We shall thus adopt the NV EOS \citep{Negele73} as that for the inner crust and the BPS EOS  \citep{Baym71} for the outer crust while focusing on the EOS of the dense core in a large 3-D and 2-D parameter space. In this section, we shall first investigate the crust-core transition density and pressure using a thermal dynamical approach. Our main goal is to examine effects of the  symmetry energy, especially its curvature $K_{\rm{sym}}$ on the transition point. We will then discuss how we construct the EOS for the core. For the purposes of this work, it is sufficient to use the simplest model for non-rotating and charge-neutral neutron stars  consisting of only $npe\mu$ particles at $\beta$-equilibrium. As we mentioned earlier, essentially all available many-body theories using various interactions have been used to predict both the EOS of SNM and symmetry energy from low to supra-saturation densities. Given their widely different predictions especially at supra-saturation densities, we try to extract parameters charactering the EOS of dense neutron-rich matter from the astrophysical observations using as little as possible predictions of any particular many-body theory and/or interaction. However, we must ensure that the EOS parameters asymptotically become naturally the ones extracted from terrestrial nuclear experiments near $\rho_0$. 

To avoid misleading the reader, here we discuss in more detail the relationship between the {\it Taylor expansions} of Eqs. (\ref{PAEb}) to (\ref{Esympara}) used to characterize the EOS near the saturation density $\rho_0$ 
in both nuclear theory and in analyzing terrestrial experiments and the {\it parameterizations} that we shall use in modeling the EOS of neutron star matter. We also address the question why we can't just adopt the widely used {\it isospin-independent} multi-parameters polytropic EOSs, see, e.g., refs \citep{Topper64,Butterworth76,Read09}, for the purposes of our work here.  Near $\rho_0$ and $\delta=0$, any nuclear energy density functional $E(\rho,\delta)$ can be Taylor expanded using Eqs. (\ref{PAEb}) to (\ref{Esympara}). The relevant expansion coefficients are determined by the $E(\rho,\delta)$ via $E_{\rm{sym}}(\rho) = \frac{1}{2}[\partial^2 E_b(\rho,\delta)/\partial\delta^2]_{\delta=0}$ for the symmetry energy, $L=3\rho_0[\partial E_{\rm{sym}}(\rho)/\partial\rho]|_{\rho=\rho_0}$ for the slope parameter of $E_{\rm{sym}}(\rho)$, $K_0=9\rho_0^2[\partial^2 E_0(\rho)/\partial\rho^2]|_{\rho=\rho_0}$ and $K_{\rm{sym}}=9\rho_0^2[\partial^2 E_{\rm{sym}}(\rho)/\partial\rho^2]|_{\rho=\rho_0}$ for the incompressibility of SNM and the curvature of $E_{\rm{sym}}(\rho)$, as well as $J_0=27\rho_0^3[\partial^3 E_0(\rho)/\partial\rho^3]|_{\rho=\rho_0}$ and $J_{\rm{sym}}=27\rho_0^3[\partial^3 E_{\rm{sym}}(\rho)/\partial\rho^3]|_{\rho=\rho_0}$ for the skewness of $E_0(\rho)$ and $E_{\rm{sym}}(\rho)$, respectively. The Taylor expansions become progressively inaccurate for large densities and do not converge when $\rho>1.5\rho_0$.  Therefore, for describing neutron star matter, we parameterize the $E_{\rm{sym}}(\rho)$ and $E_0(\rho)$ in exactly the same form as in Eqs. (\ref{PAEb}) to (\ref{Esympara}) but their coefficients are no longer given by the above expressions from any energy density functional $E(\rho,\delta)$. Instead, we treat them as unknown parameters to be extracted from astrophysical observations. More specifically, while the NV+BPS EOSs are used for the crust, starting from the crust-core transition density we parameterize the $E_0(\rho)$ and $E_{\rm{sym}}(\rho)$ as
\begin{equation}\label{E0p}
  E_{0}(\rho)=E_0(\rho_0)+\frac{K^{\rm{NS}}}{18}(\frac{\rho}{\rho_0}-1)^2+\frac{J^{\rm{NS}}}{162}(\frac{\rho}{\rho_0}-1)^3,
\end{equation}
\begin{equation}\label{Esymp}
  E_{\rm{sym}}(\rho)=E_{\rm{sym}}(\rho_0)+\frac{L^{\rm{NS}}}{3}(\frac{\rho}{\rho_0}-1)+\frac{K^{\rm{NS}}_{\rm{sym}}}{18}(\frac{\rho}{\rho_0}-1)^2+\frac{J^{\rm{NS}}_{\rm{sym}}}{162}(\frac{\rho}{\rho_0}-1)^3
\end{equation} 
using the parameters $K^{\rm{NS}}, J^{\rm{NS}}, L^{\rm{NS}}, K^{\rm{NS}}_{\rm{sym}}$ and $J^{\rm{NS}}_{\rm{sym}}$. These {\it parameterizations} naturally become the Taylor expansions in the limit of $\rho\rightarrow \rho_0$. As parameterizations, mathematically they can be used at any density without the convergence issue associated with the Taylor expansions.
Any parameterized EOS has to satisfy all the constraints (asymptotic boundary conditions) for the EOS near $\rho_0$ from terrestrial experiments.  The above parameterizations facilitate comparisons with the terrestrial constraints around $\rho_0$. When the above parameterizations are used at high densities for modeling the core EOS of neutron stars, the parameters in the above two equations are not necessarily to be the same as the Taylor expansion coefficients in Eqs. (\ref{PAEb}) to (\ref{Esympara}). While both mathematically and physically, the $L$ and $L^{\rm{NS}}$ are expected to be very close, the high-order coefficients (parameters) in the Taylor expansions (parameterizations) may be very different. Unfortunately, very little is known about these parameters from experiments/observations and the model predictions for the EOS of dense neutron-rich matter diverge. 
Since the uncertainty ranges of the Taylor expansion coefficients $L, K_{\rm{sym}}$, $J_{\rm{sym}}$ and $J_0$ are already so large, it is then meaningful to use their uncertainty ranges as the starting/reference ranges in our search for the high-density EOS parameters $K^{\rm{NS}}, J^{\rm{NS}}, L^{\rm{NS}}, K^{\rm{NS}}_{\rm{sym}}$ and $J^{\rm{NS}}_{\rm{sym}}$ from studying properties of neutron stars.  

With the above definitions and explanations, we now introduce/discuss a convention/simplification that we shall use in the following. It is consistent with the existing convention widely used in the literature for the Eq (\ref{PAEb}). The latter is often referred as the empirical parabolic law/approximation of the EOS of isospin asymmetric nuclear matter \citep{Bom91}. Namely, it is  well known in nuclear physics that the Eq (\ref{PAEb}) has the dual meanings of being a Taylor expansion in $\delta$ in the limit of $\delta\rightarrow 0$ on one hand, and on the other hand being a parameterization when it is used in very neutron-rich matter where $\delta\rightarrow 1$. Adopting this convention and be consistent with that used for the Eq (\ref{PAEb}), albeit confusing without the above explanations, it is really not necessary to write both the Taylor expansions of  Eqs. (\ref{PAEb}) to (\ref{Esympara}) and the parameterizations of Eqs. (\ref{E0p}) and (\ref{Esymp}) as they have the same form although different meanings for the coefficients/parameters involved. In the following discussions,  when the Eqs. (\ref{PAEb}) to (\ref{Esympara}) are used in describing terrestrial nuclear EOS or predictions of nuclear energy density functional theories, they are Taylor expansions near $\rho_0$ and $\delta=0$. On the other hand, when they are used in describing the EOS of neutron stars, they are parameterizations to be constrained by astrophysical observations. Then, all three equations in Eqs. (\ref{PAEb}) to (\ref{Esympara}) have the dual meanings in describing the EOS of isospin asymmetric matter encountered in both terrestrial experiments and neutron stars within broad ranges of $\delta$ and $\rho$ without the convergence problem. 

In the sense that because there is no reliable theory for the EOS of neutron-rich matter significantly above the saturation density one has to use parameterizations to describe the EOS of neutron star matter,  
the spirit of the parameterizations of Eqs. (\ref{E0p}) and (\ref{Esymp}) is essentially the same as many other parameterizations used in the literature, see, e.g., refs. \citet{Gandolfi09,Gandolfi12,Steiner16}. However, instead of requiring indirectly the involved parameters to reproduce the known properties of symmetric nuclear matter and the symmetry energy near $\rho_0$ through sometimes complicated equations, we use directly 
the coefficients of the Eqs. (\ref{E0para}) and (\ref{Esympara}). These coefficients as the asymptotic boundary conditions of the EOS are themselves known near $\rho_0$ and $\delta=0$ either through experiments or converged predictions of many reliable models, see, e.g., refs. \citep{Danielewicz02,Shlomo06,Lwchen09,Steiner10,Piekarewicz10,Khan12,Dutra12,Dutra14,Li13,Colo14,Cai14,Zhang17}.
In fact, parameterizations similar to the Eqs. (\ref{E0para}) and (\ref{Esympara}), albeit often truncated at the first order in density for $E_{\rm{sym}}(\rho)$ and second order for $E_0(\rho)$, i.e., using $L$ and $K_0$ only, have already been used successfully in studying various properties of neutron stars, see, e.g., refs.\citep{Oyamatsu07,Sotani12}. Also, a similar approach of describing approximately the EOS of dense neutron-rich matter was recently proposed in ref. \citep{MM1} and successfully used in studying properties of both neutron stars \citep{MM2} and finite nuclei \citep{MM3} with Bayesian perspectives.
With the above cautions, conventions and justifications, we shall use the parameterized EOS for the core of neutron stars as described in more detail in section \ref{sec2.3}.

The multi-parameters polytropic EOSs, see, e.g., refs \citep{Topper64,Butterworth76,Read09} are widely used in modeling the core EOS of neutron stars at supra-saturation densities. Why do not we simply follow this popular approach? This is mainly because the polytropes used so far depend only on the nucleon number (energy) density but explicitly independent of the isospin asymmetry $\delta$. To extract information about the high-density symmetry energy from neutron stars, we need to know explicitly the underlying isospin composition of the pressure/density. For example, to our best knowledge, all Bayesian analyses done so far using masses and radii of neutron stars have adopted the polytropes. They infer only the total pressures at a few fiducial high densities without any information about the isospin content of the matter at those densities. Indeed, to invert mathematically the TOV equation one only needs to know the total pressure as a function of density regardless of its microphysics composition. However, to understand the entire structure, composition and cooling mechanism of  neutron stars, it is necessary to know the isospin dependence of its EOS from the crust to the dense core consistently. Since the pressure has a term proportional to $\delta^2\cdot dE_{\rm{sym}}(\rho)/d\rho$ where the isospin-asymmetry profile $\delta(\rho)$ in neutron stars at $\beta$-equilibrium is determined uniquely by the $E_{\rm{sym}}(\rho )$ as we shall discuss in more detail below, by directly parameterizing the $E_0(\rho)$ and $E_{\rm{sym}}(\rho )$, separately, and then evaluate the resulting pressure as a function of both $\rho$ and $\delta (\rho)$, we shall obtain information about the high-density $E_{\rm{sym}}(\rho )$ underlying the pressure-density relation $P(\rho)$. While directly parameterizing the $P(\rho)$ do not facilitate the extraction of such information about the high-density symmetry energy from astrophysical observations. 

\subsection{Sampling the density dependence of nuclear symmetry energy}
\label{sec2.1}
\begin{figure}[ht]
\begin{center}
  \includegraphics[width=5cm]{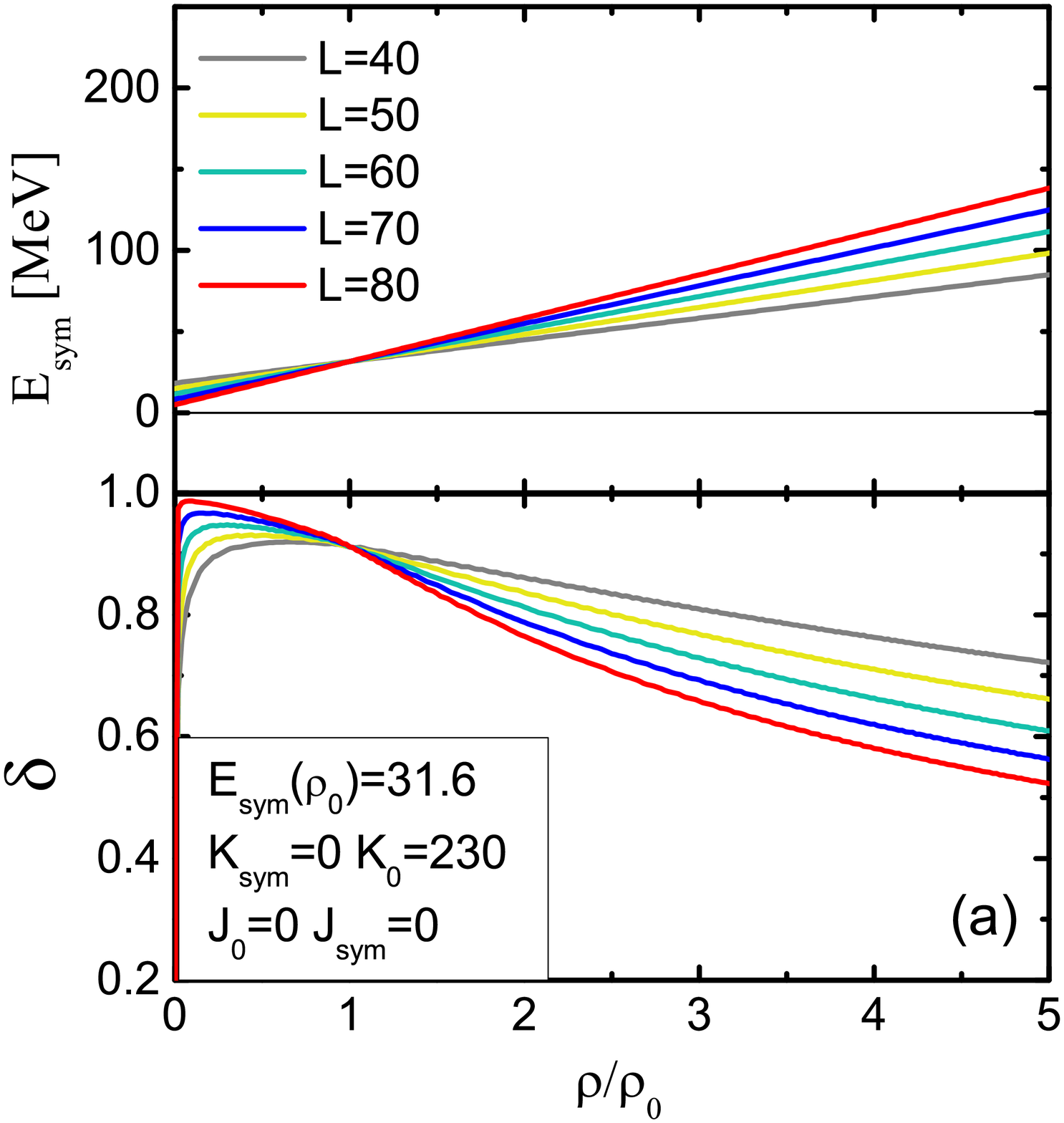}
  \includegraphics[width=5cm]{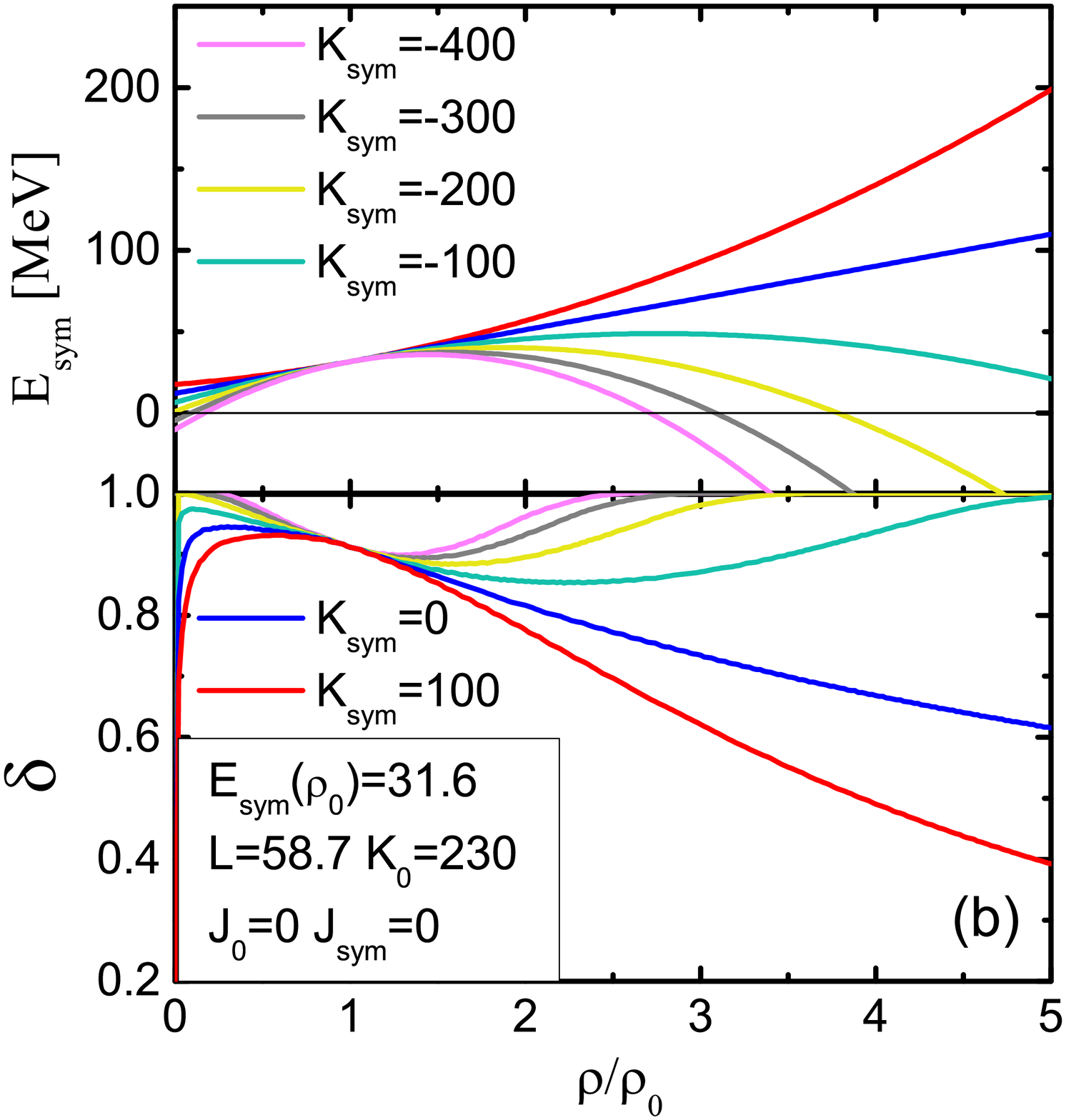}
  \includegraphics[width=5cm]{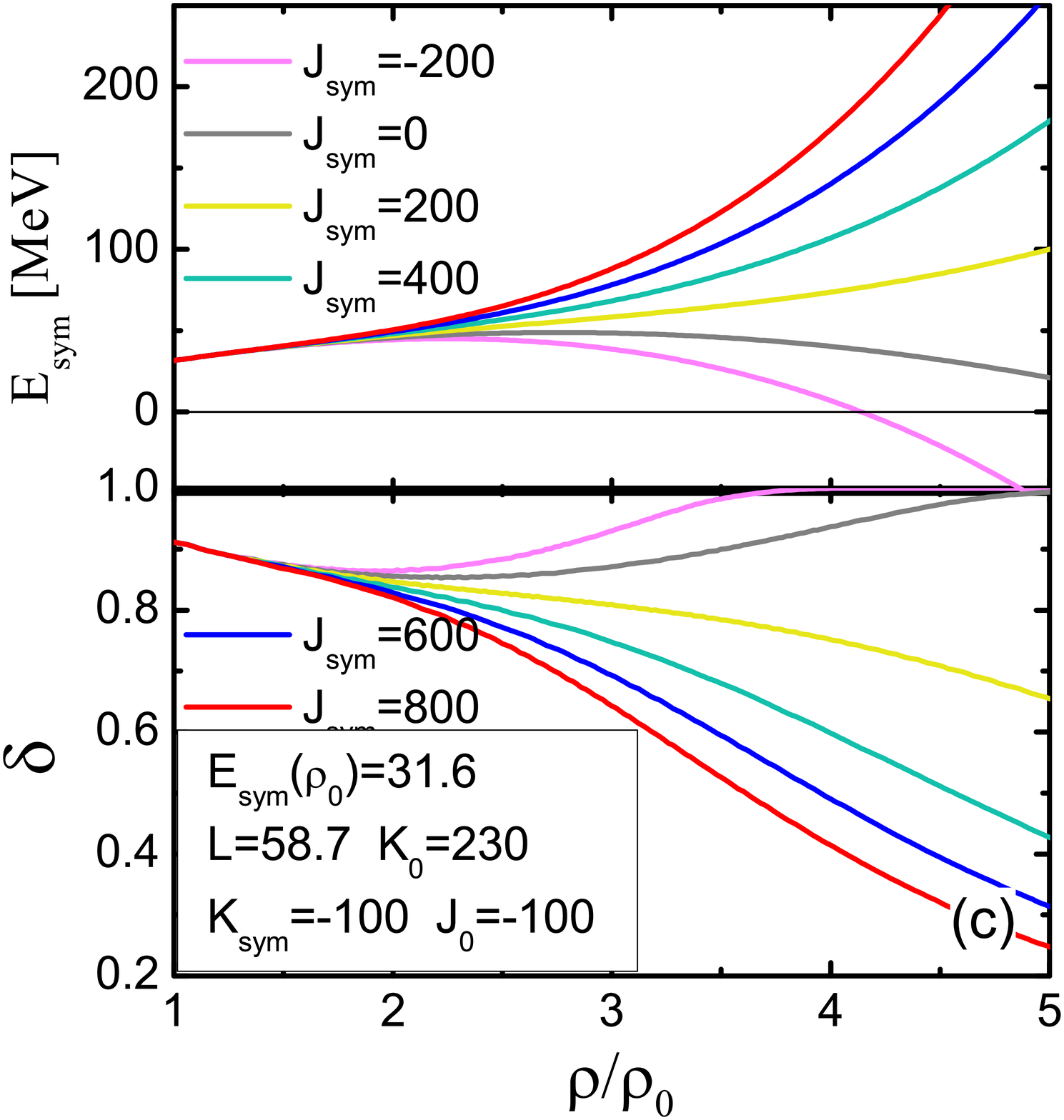}
  \caption{(color online) The symmetry energy $E_{\rm{sym}}(\rho)$ and isospin asymmetry $\delta(\rho)$ in neutron star matter at $\beta$-equilibrium as a function of the reduced density $\rho/\rho_0$ for $L=40$, 50, 60, 70, and 80 MeV (a), $K_{\rm{sym}}=-400$, -300, -200, -100, 0, and 100 MeV (b), and $J_{\rm{sym}}=-200$, 0, 200, and 400 MeV (c), respectively. All parameters are in unit of MeV. }\label{Ksymeffect}
\end{center}
\end{figure}
First of all, we illustrate the broad variation of $E_{\rm{sym}}(\rho)$ by varying independently the coefficients in Eq. (\ref{Esympara}) within their known uncertainty ranges. It is well known that the isospin asymmetry $\delta(\rho)$ in neutron stars at a given baryon density $\rho$ is uniquely determined by the $E_{\rm{sym}}(\rho)$ in Eq. (\ref{PAEb}) through the charge neutrality and $\beta$ equilibrium conditions as we shall recall more formally in section \ref{sec2.3}. Generally speaking, because of the $E_{\rm{sym}}(\rho)\cdot \delta^2$ term in the EOS, a higher value of $E_{\rm{sym}}(\rho)$ will lead to a smaller $\delta(\rho)$ at $\beta$ equilibrium.
As quantitative examples,  shown in Figure \ref{Ksymeffect} are the $E_{\rm{sym}}(\rho)$ and $\delta(\rho)$ as functions of reduced density by varying only one coefficient each time while fixing all others: (a) $L=40$, 50, 60, 70, and 80 MeV, (b) $K_{\rm{sym}}=-400$, -300, -200, -100, 0, and 100 MeV, and (c) $J_{\rm{sym}}=-200$, 0, 200, and 400 MeV.  As their names indicate, the slope $L$, curvature $K_{\rm{sym}}$ and skewness  $J_{\rm{sym}}$ of symmetry energy play different roles and in order become increasingly more important at higher densities. Obviously, variations of them within their currently known uncertainty ranges allow us to sample very different behaviors of the $E_{\rm{sym}}(\rho)$ and the corresponding $\delta(\rho)$.

It is worth noting that some combinations of the parameters lead to a decreasing $E_{\rm{sym}}(\rho)$ that may even become negative at high densities. As summarized earlier in \citet{Kut06} and reviewed very recently in \citet{Li18}, such kind of super-soft $E_{\rm{sym}}(\rho)$ at high densities was predicted in a number of theoretical calculations using various interactions.  At very high densities, when the short-range repulsive tensor force due to the $\rho$-meson exchange makes the EOS of SNM increase faster with density than that of pure neutron matter where the tensor force is much weaker, the $E_{\rm{sym}}(\rho)$ decreases or even becomes negative at high densities \citep{Pan72,WFF,LCK08}. To our best knowledge, such a seemingly unusual high-density behavior of the $E_{\rm{sym}}(\rho)$ is not excluded by neither any known fundamental physics principle nor experiments/observations so far. 
Possible consequences of such kind of symmetry energies are discussed in more detail in refs. \citep{Kut06,Li18} and references therein. 
In fact, EOSs with a super-soft $E_{\rm{sym}}(\rho)$ can still support massive neutron stars if the SNM parts of the EOSs are sufficiently stiff even without the help from the new light mesons proposed and/or possible modified strong-field gravity for massive objects, see, e.g., refs. \citep{Kri09,Wen09,Wlin14,Jiang15}. Interestingly, while not completely settled yet \citep{Li17}, there are indeed some circumstantial evidences from intermediate-relativistic energy heavy-ion collisions indicating that the \esym may become super-soft above about $2\rho_0$ \citep{XiaoPRL}. Currently, devoted efforts are being made by the intermediate-relativistic heavy-ion reaction community to pin down the high-density behavior of nuclear symmetry energy, see, e.g., refs. \citep{Tesym,ASY-EOS,Xu16-t,Tsang17,Herman17}.

As we shall discuss in detail next, the $E_{\rm{sym}}(\rho)$ around the crust-core transition is mostly controlled by the $L$ and $K_{\rm{sym}}$ parameters when the $E_{\rm{sym}}(\rho_0)$ is fixed at its most probable empirical value of 31.6 MeV. The variation of $L$ from 40 to 80 MeV and $K_{\rm{sym}}$ from -400 to 100 MeV allows us to sample the usual behavior of $E_{\rm{sym}}(\rho_0)$ predicted by various nuclear many-body theories in the sub-saturation density region and within their known empirical constraints at $\rho_0$.

\subsection{Imprints of the density dependence of nuclear symmetry energy on the crust-core transition in neutron stars}
\label{sec2.2}
Although the crust of neutron stars contributes only a small fraction of the total mass and radius, it plays an important role in various astrophysical phenomena \citep{Baym71,Baym71b,Pethick95,Link99,Lattimer00,Lattimer01,Steiner05,Lattimer07,Chamel08,Sotani12,Newton13,Pons13,Piekarewicz14,Horowitz15}. Critical to many effects of the crust is the transition density $\rho_t$ between the inner crust and the outer core of neutron stars. Previous studies have found consistently that the transition density is very sensitive to the density dependence of nuclear symmetry energy. In particular, the role of the slope parameter $L$ has been extensively studied, see, e.g., refs. \citep{Douchin00,Kubis04,Kubis07,Lattimer07,Oyamatsu07}. Often the studies employ predictions of a particular nuclear many-body theory where the values of $L$ and $K_{\rm{sym}}$ are normally correlated. Here we shall first study the individual roles of the $L$ and $K_{\rm{sym}}$ in determining the core-crust transition properties, then contours of the transition density and pressure in the $L$ versus $K_{\rm{sym}}$ plane. Finally, effects of the $L-K_{\rm{sym}}$ correlation based on the systematics from analyzing over 500 nuclear energy density functions are examined.

\begin{figure}[ht]
\begin{center}
  \includegraphics[width=8cm]{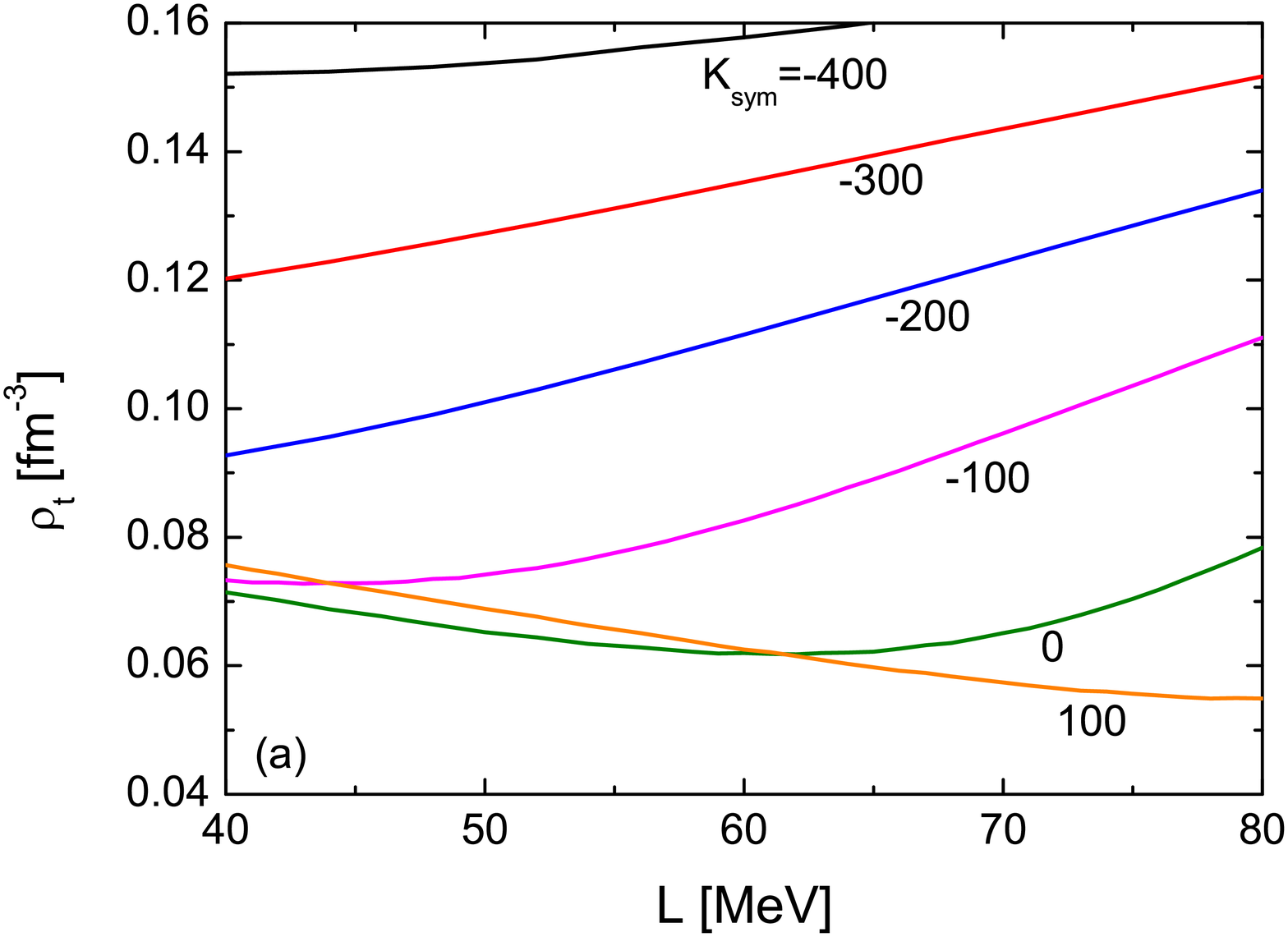}
  \includegraphics[width=8cm]{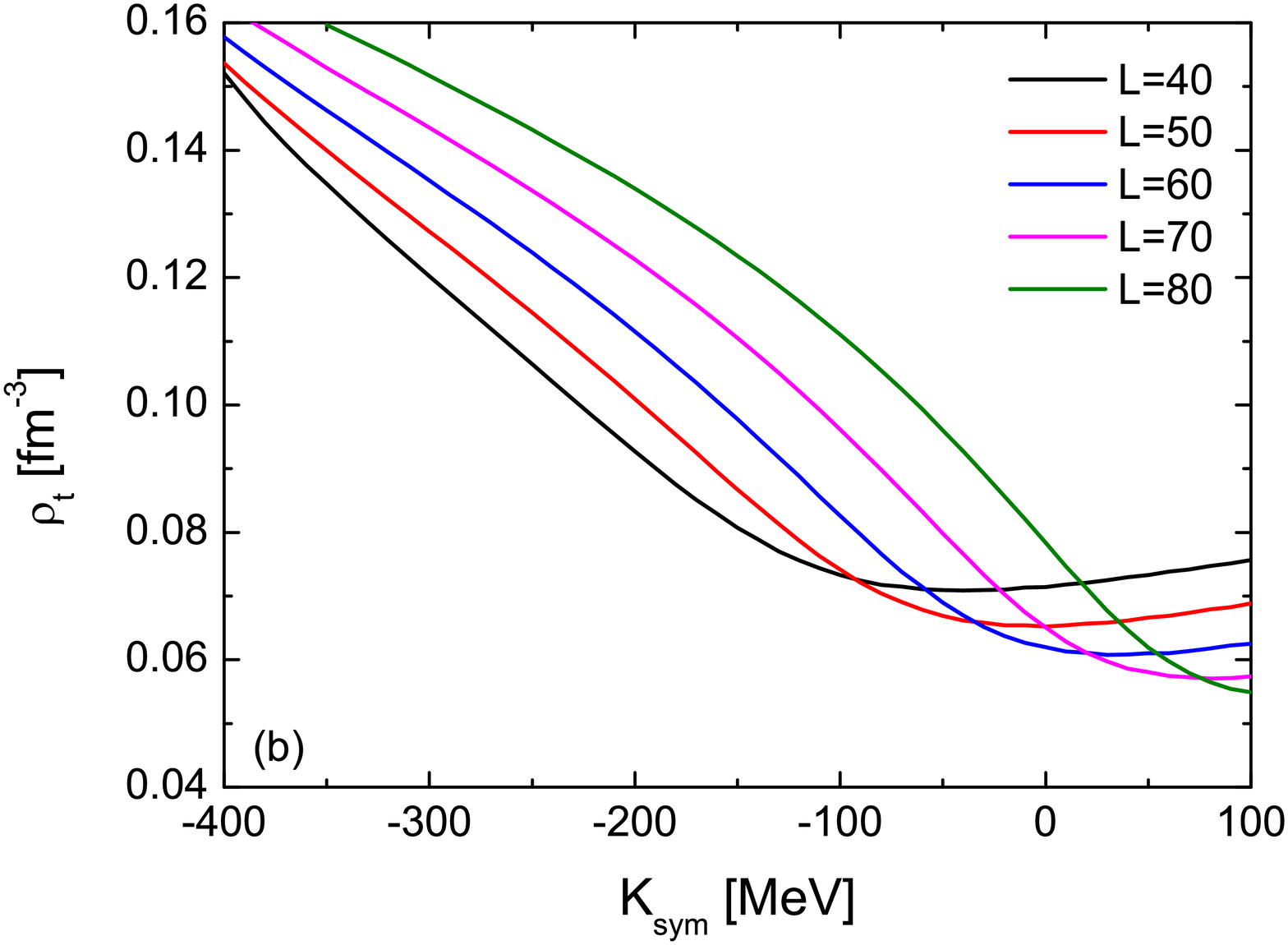}
  \caption{(color online) The crust-core transition density $\rho_t$ as a function of $L$ (left panel) with $K_{\rm{sym}}$ fixed at -400, -300, -200, -100, 0, and 100 MeV, and $K_{\rm{sym}}$ (right panel) with $L$ fixed at 40, 50, 60, 70, and 80 MeV, respectively.}\label{LKsymfix}
\end{center}
\end{figure}
In the present study, we employ the thermodynamical approach \citep{Kubis04,Kubis07,Lattimer07} to estimate the crust-core transition point where the uniform matter becomes unstable against being separated into a mixture of single nucleons and their clusters. The method is known to slightly overestimate the transition density compared to the dynamical approach \citep{Xu09,Ducoin11,Pro14} but sufficiently good for the purposes of this work. Specifically, the transition density is determined by the vanishing effective incompressibility of neutron star matter at $\beta$ equilibrium under the charge neutrality condition \citep{Kubis04,Kubis07,Lattimer07}, i.e.,
\begin{equation}\label{tPA}
K_\mu=\rho^2\frac{d^2E_0}{d\rho^2}+2\rho\frac{dE_0}{d\rho}+\delta^2\left[\rho^2\frac{d^2E_{\rm{sym}}}{d\rho^2}+2\rho\frac{dE_{\rm{sym}}}{d\rho}-2E^{-1}_{\rm{sym}}(\rho\frac{dE_{\rm{sym}}}{d\rho})^2\right]=0.
\end{equation}
This approach has been used extensively in the literature to locate the transition point using various EOSs \citep[see, e.g., refs.][]{Xu09,Ducoin11,Pro14,Routray16}.  Enclosed in the bracket of the above expression for $K_\mu$ are the first-order and second-order derivatives of the symmetry energy, i.e., quantities directly determining the $L$ and  $K_{\rm{sym}}$. It is thus necessary and interesting to first explore separate roles of the latter on the transition density. Shown in Figure \ref{LKsymfix} are the transition density $\rho_t$ as functions of $L$ with different values of $K_{\rm{sym}}$ in the window-a and $K_{\rm{sym}}$ with different values of $L$ in the window-b, respectively. It is clearly seen that the $\rho_t$ changes more dramatically with the variation of $K_{\rm{sym}}$ than $L$ in their respective uncertainty ranges. This is mainly because the last two terms in the expression for $K_\mu$ largely cancel out, leaving the curvature of $E_{\rm{sym}}(\rho)$ dominate. In addition, the value of $L$ is already relatively well constrained in a smaller range than the $K_{\rm{sym}}$, making the variation of $\rho_t$ with $L$ look weaker.

\begin{figure}[ht]
\begin{center}
  \includegraphics[width=8cm]{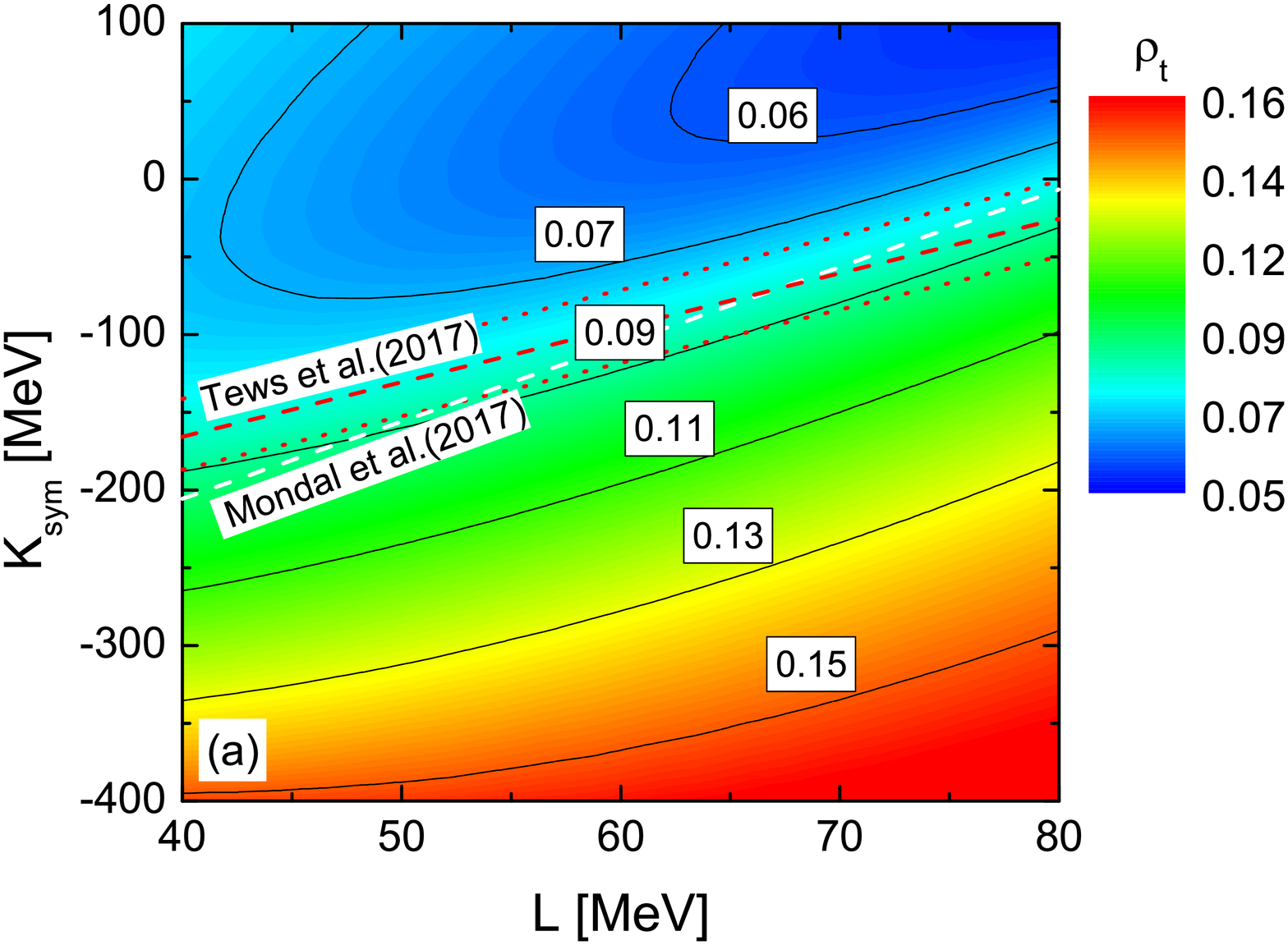}
  \includegraphics[width=8cm]{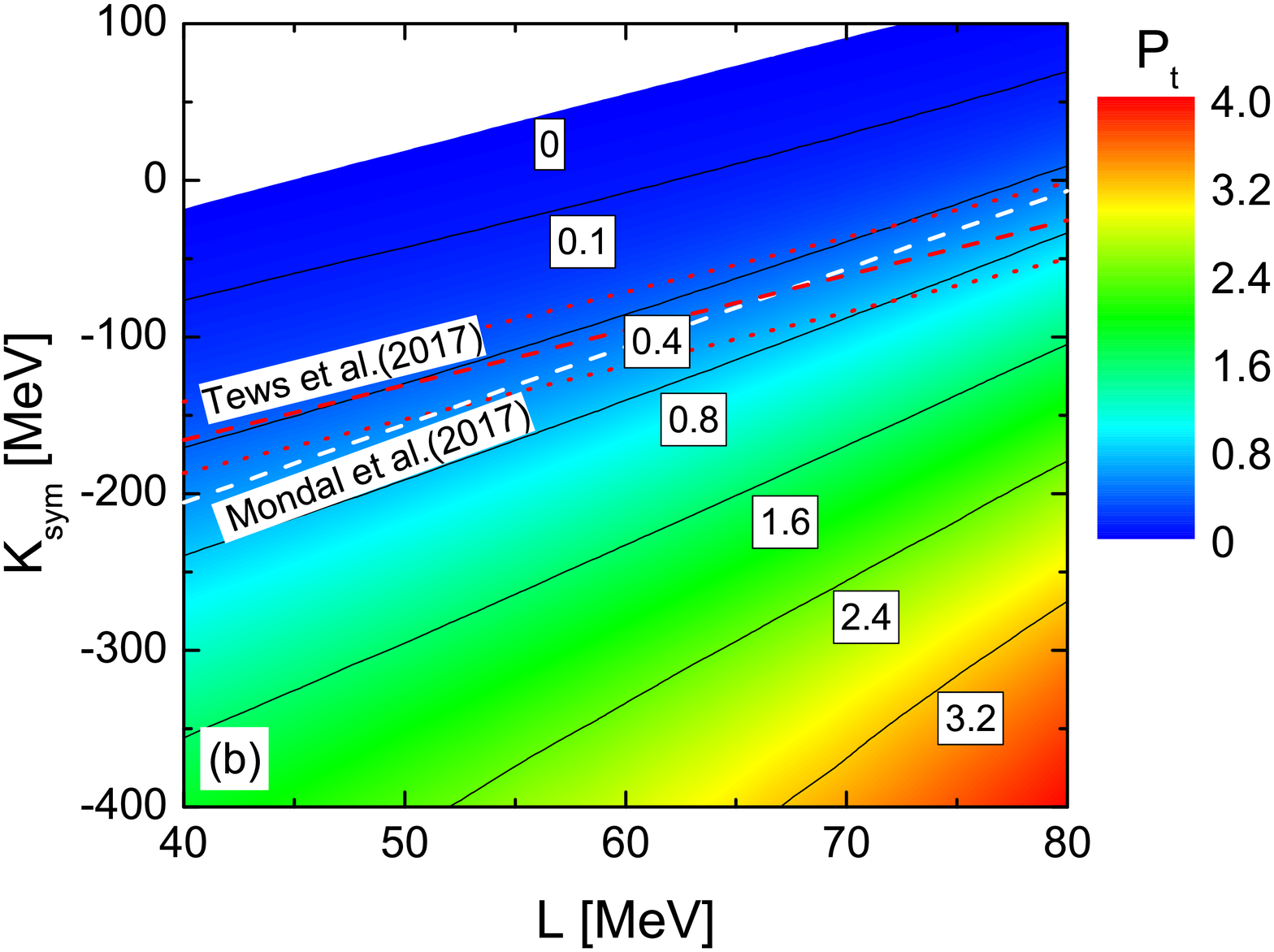}
  \caption{(color online) Contours of the crust-core transition density $\rho_t$ in fm$^{-3}$ (a) and the corresponding pressure $P_t$ in MeV fm$^{-3}$ (b) in the $L-K_{\rm{sym}}$ plane. Lines with fixed values of transition densities and pressures are labeled. The white and red dashed lines are the correlations between $K_{\rm{sym}}$ and $L$ from \citet{Tews17} and \citet{Mondal17} (see text), respectively. The white region in (b) is where the transition pressure vanishes.} \label{LKsymrhoP}
\end{center}
\end{figure}

Next, we examine the transition density $\rho_t$ and pressure $P_t$ by varying both the $K_{\rm{sym}}$ and $L$ within their uncertainty ranges continuously. Shown in the two windows of Figure \ref{LKsymrhoP} are contours of constant transition densities $\rho_t$ and pressures $P_t$ in the $L-K_{\rm{sym}}$ plane, respectively. For transition densities larger than $\rho_t=0.07$ fm$^{-3}$, the required $K_{\rm{sym}}$ increases monotonically with $L$, while different behaviors are observed for lower values of $\rho_t$. The lowest transition density about $\rho_t=0.0549$ fm$^{-3}$ appears around the boundary corner at $L=77$ MeV and $K_{\rm{sym}}=100$ MeV. Different from the contours of constant $\rho_t$, for a fixed $P_t$,  the required $K_{\rm{sym}}$ always increases linearly with $L$ before reaching the $P_t=0$ boundary along the line $K_{\rm{sym}}=3.64 L-163.96~(\rm{MeV})$. The latter is used as 
a limit in exploring properties of neutron stars in the EOS parameter space.

As indicated earlier, in using the Eqs. (\ref{E0para}) and (\ref{Esympara}) we assume the coefficients are independent and intend to
constrain them directly from observations using as little as possible predictions of any particular many-body theory. Thus, in determining the crust-core transition point for constructing the EOS of neutron stars, we freely vary the
$K_{\rm{sym}}$ and $L$ within their uncertainty ranges specified earlier. Nevertheless, theoretically predicted values of the $K_{\rm{sym}}$ and $L$ are often correlated when model ingredients and/or interactions are varied.
Thus, for a consistency check we also study how the predicted correlation between $K_{\rm{sym}}$ and $L$ may limit the transition point.
As discussed by many people in the literature \citep[see, e.g., refs.][]{Far78,Ducoin11,Pro14,Pearson14,Danielewicz09,Lwchen09,Vidana09,Ducoin11,Mondal17,Tews17},
based on the systematics of many predictions using various many-body theories and interactions, an approximately universal and linear correlation exists between the $K_{\rm{sym}}$ and $L$.
For example, using totally over 500 energy density functions including 263 Relativistic Mean Field (RMF) models, Hartree-Fok calculations using 240 Skyrme \citep{Dutra12,Dutra14} as well as some realistic and Gogny interactions, \citet{Mondal17} found the following $K_{\rm{sym}}-L$ correlation
\begin{equation}\label{KsynLrelationship1}
  K_{\rm{sym}}=(-4.97\pm0.07)(3E_{\rm{sym}}(\rho_0)-L)+66.8\pm2.14 ~~\rm{MeV}.
\end{equation}
Using essentially the same sets of energy density functionals but requiring $0.149<\rho_0<0.17$ fm$^{-3}$, $-17<E_0(\rho_0)<-15$ MeV, $25<E_{\rm{sym}}(\rho_0)<36$ MeV, and $180<K_0<275$ MeV, \citet{Tews17} rejected some of the energy functionals. They found the following $K_{\rm{sym}}-L$ correlation using the remaining 188 Skyrme and 73 RMF interactions
\begin{equation}\label{KsynLrelationship2}
  K_{\rm{sym}}=3.501L-305.67\pm24.26~(\pm56.59) ~~\rm{MeV},
\end{equation}
where the $\pm24.26$ and $\pm56.59$ are error bars including 68.3\% and 95.4\% of the accepted EOSs, respectively.
 In Figure \ref{LKsymrhoP}, the above two $K_{\rm{sym}}-L$ correlation functions are shown with the white and red dashed lines, separately. The 68.3\% uncertainty range ($\pm24.26$ MeV) in Eq. (\ref{KsynLrelationship2}) is indicated by the red dotted lines while that of Eq. (\ref{KsynLrelationship1}) is too small to be shown here. While the parameterizations in Eqs. (\ref{KsynLrelationship1}) and (\ref{KsynLrelationship2}) are consistent, their mean values  have slightly different slopes because of the different selection criteria used. Nevertheless, if the uncertainty range of Eq. (\ref{KsynLrelationship2}) is enlarged from 68.3\% to 95.4\% of the accepted EOSs, the parameterization in Eq. (\ref{KsynLrelationship1}) can then be fully covered by that in Eq. (\ref{KsynLrelationship2}).  Using the above two correlations, the transition density and pressure are then restricted to be about $\rho_t=0.08$ fm$^{-3}$ and $P_t=0.40$ MeV fm$^{-3}$, respectively. These values are consistent with the crust-core transition properties often used in the literature. To this end, especially since some of the apparent correlations among EOS parameters from model calculations may be spurious \citep{MM1},  it is worth noting that while the $K_{\rm{sym}}-L$ correlation from the systematics of over 500 energy density functions is very useful for the consistency check, it is still necessary and important to determine the individual values of $K_{\rm{sym}}$ and $L$ from experiments/observations. In our following calculations, we thus use consistently the crust-core transition density and pressure by varying independently the  $K_{\rm{sym}}$ and $L$ values within their respective uncertain ranges without using any of the about two correlation functions.

\subsection{The core EOS of neutron stars}
\label{sec2.3}
For completeness and the ease of our discussions in the following, we first recall here the formalism for calculating the EOS in the cores of neutron stars.
The total energy density $\epsilon(\rho, \delta)$ of charge neutral $npe\mu$ matter at $\beta$-equilibrium can be written as
\begin{equation}\label{energydensity}
  \epsilon(\rho, \delta)=\epsilon_b(\rho, \delta)+\epsilon_l(\rho, \delta),
\end{equation}
where $\epsilon_b(\rho, \delta)$ and $\epsilon_l(\rho, \delta)$ are the energy density of baryons and leptons, respectively. The $\epsilon_b(\rho, \delta)$ can be calculated from
\begin{eqnarray}\label{Ebpa}
  \epsilon_b(\rho,\delta)=\rho E(\rho,\delta)+\rho M_N,
\end{eqnarray}
where the specific energy $E(\rho,\delta)$ of baryons is given in Eq. (\ref{PAEb}) and $M_N$ is the average rest mass of nucleons. The $\epsilon_l(\rho, \delta)$ from the noninteracting Fermi gas model can be expressed as ($\hbar=c=1$) \citep{Oppenheimer39}
\begin{equation}
  \epsilon_l(\rho, \delta)=\eta\phi(t)
\end{equation}
with
\begin{equation}
  \eta=\frac{m_l^4}{8\pi^2},~~
  \phi(t)=t\sqrt{1+t^2}(1+2t^2)-ln(t+\sqrt{1+t^2}),
\end{equation}
and
\begin{equation}
  t=\frac{(3\pi^2\rho_l)^{1/3}}{m_l}.
\end{equation}
The chemical potential of particle $i$ can be calculated from
\begin{equation}\label{chemicalnpe}
  \mu_i=\frac{\partial\epsilon(\rho,\delta)}{\partial\rho_i}.
\end{equation}
The isospin asymmetry $\delta(\rho)$ and relative particle fractions at different densities in neutron stars are obtained through the $\beta$-equilibrium condition
$
  \mu_n-\mu_p=\mu_e=\mu_\mu\approx4\delta E_{\rm{sym}}(\rho)
$
and the charge neutrality condition
$  \rho_p=\rho_e+\rho_\mu.
$
The pressure of the system can be calculated numerically from
\begin{equation}\label{pressure}
  P(\rho, \delta)=\rho^2\frac{d\epsilon(\rho,\delta)/\rho}{d\rho}.
\end{equation}
The above expressions allow us to calculate the core EOS which is connected smoothly at the core-crust transition point to the NV EOS \citep{Negele73} for the inner crust
followed by the BPS EOS  \citep{Baym71} for the outer crust. 

As mentioned earlier, the EOS parameters $K_0$, $E_{\rm{sym}}(\rho_0)$ and $L$ near $\rho_0$ are relatively well determined. To investigate how/what high-density EOS parameters are constrained by the three astrophysical observations considered in this work, we construct the EOS of neutron star matter by varying the poorly known $J_0$, $K_{\rm{sym}}$ and $J_{\rm{sym}}$ characterizing the EOS of dense neutron-rich nucleonic matter. In principle, all coefficients used in Eqs. (\ref{E0para}) and (\ref{Esympara}) should be varied simultaneously within a multivariant Bayesian inference \citep{Steiner10,Raithel16,MM2}. Such a study is in progress. In the present work, we shall perform traditional analyses first in the 3-D parameter space spanned by $J_0$, $K_{\rm{sym}}$ and $J_{\rm{sym}}$ while fixing all other parameters at their currently known most probable values. Equivalent to re-parameterizing the EOS of SNM and $E_{\rm{sym}}(\rho)$ with less parameters as often done in the literature, or expanding the Eqs. (\ref{E0para}) and (\ref{Esympara}) only up to $[(\rho-\rho_0)/3\rho_0]^2$, we shall also explore the EOS in the 2-D parameter space of $L$ and $K_{\rm{sym}}$ by setting  $J_0=J_{\rm{sym}}=0$ while keeping other parameters at their known most probable values. The two cases studied here are similar in spirit to using different numbers of piecewise polytropes or parameters to model the EOS of dense neutron-rich matter. Naturally, the values of the parameters involved may be different in the two cases, but they should all asymptotically approach the same existing constraints on them near $\rho_0$.

Certainly, there have been continuous efforts in both astrophysics and nuclear physics to constrain the EOS parameters in both cases. For example,  from the pressure of SNM constrained by nucleon collective flow data in relativistic heavy-ion collisions \citep{Danielewicz02}, a constraint of $-1280\leq J_0 \leq -10$ MeV was obtained by \citet{Cai14}. Combining it with the mass of neutron star PSR J0348+0432, they further narrowed it down to $-494 \leq J_0 \leq -10$ MeV. By analyzing X-ray bursts, \citet{Steiner10} extracted a value of $-690\leq J_0\leq -208$ or $-790\leq J_0\leq -330$ MeV assuming a photospheric radius of $r_{ph} \gg R$ or $r_{ph} = R$, respectively. All of these constraints on $J_0$ overlap but have different uncertainty ranges.
Similarly, the $K_{\rm{sym}}$ and $J_{\rm{sym}}$ have not been well constrained either by any experiments/observations so far.
Nevertheless, the systematics of over 500 RMF and SHF energy density functionals indicates the following range: $-800\leq J_0\leq 400$ MeV \citep{Dutra12,Dutra14}, $-400 \leq K_{\rm{sym}} \leq 100$ MeV and $-200 \leq J_{\rm{sym}} \leq 800$ MeV, respectively \citep[see, e.g.,][]{Lwchen09,Dutra12,Dutra14,Colo14,Zhang17}. Thus, we adopt these ranges for the parameters $J_0$, $K_{\rm{sym}}$ and $J_{\rm{sym}}$ to be consistent with both existing experimental and theoretical findings.

\begin{figure}[ht]
\begin{center}
  \includegraphics[width=8cm]{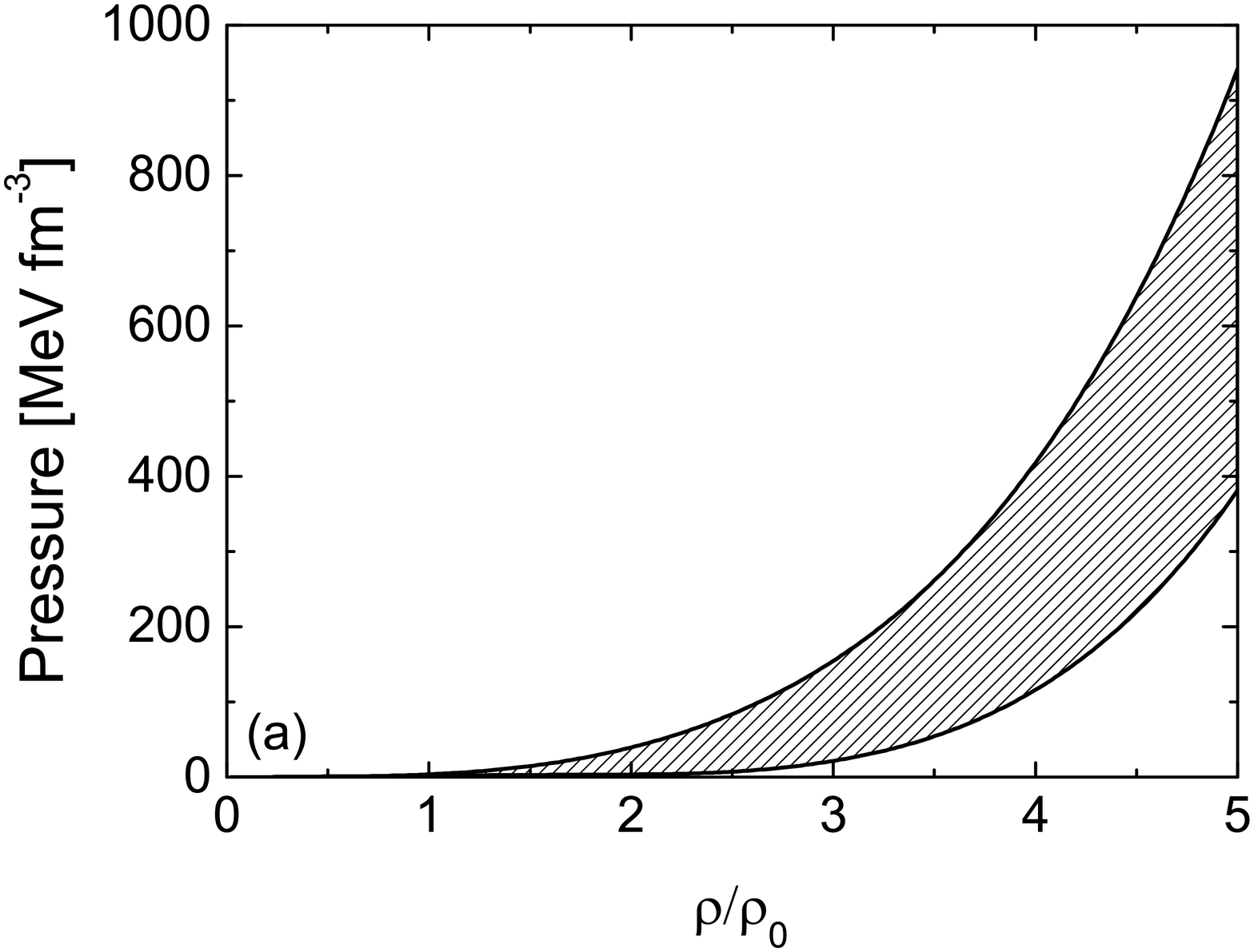}
  \includegraphics[width=8cm]{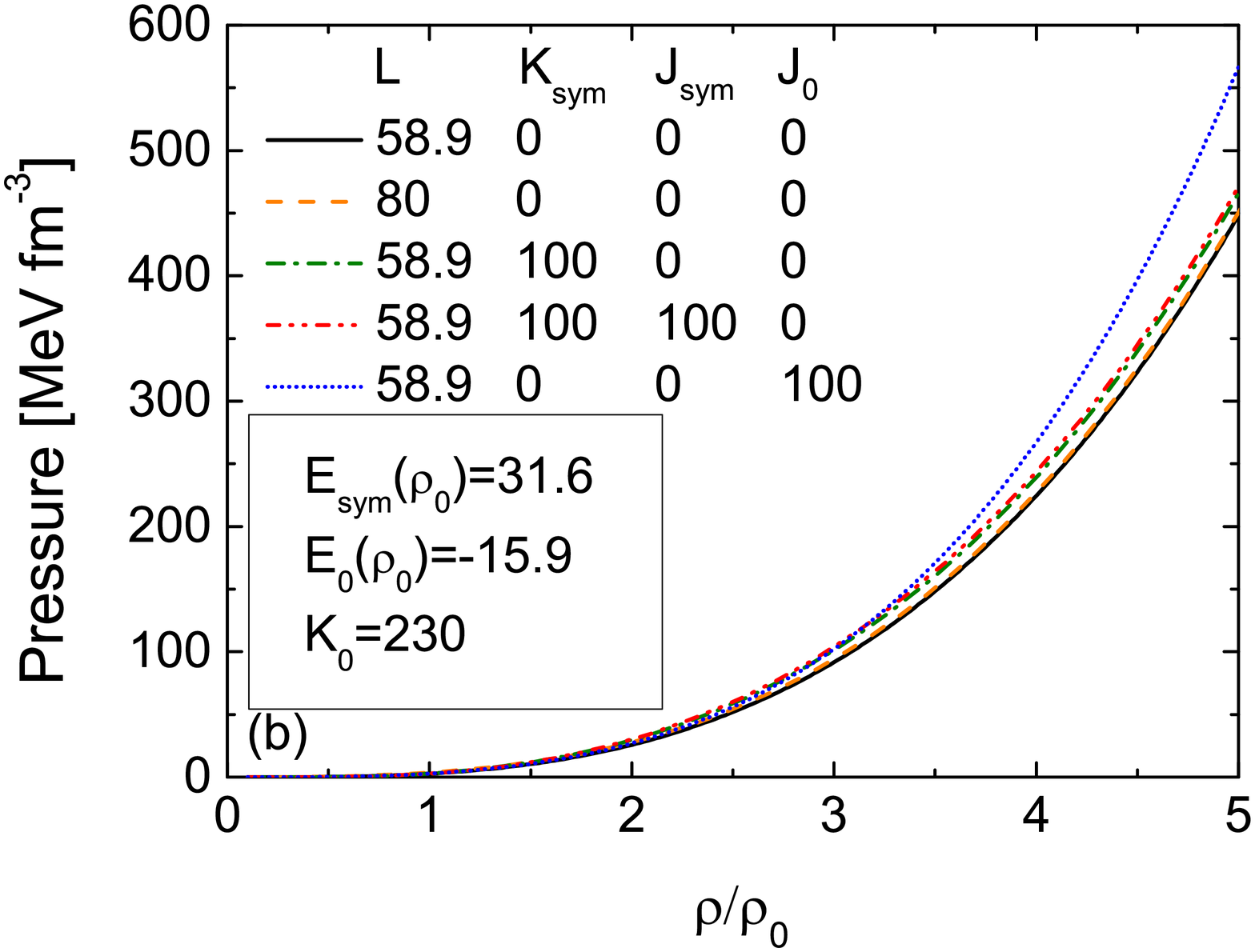}
  \caption{(color online) The pressure of neutron star matter as a function of reduced nucleon density $\rho/\rho_0$. (a): The region of pressure covered by the EOS parameters considered in the present work. (b): Effects of some EOS parameters on the pressure. All parameters are in unit of MeV.}\label{pres}
\end{center}
\end{figure}
Within the above uncertainty ranges of the EOS parameters, the pressure in neutron stars can be varied within the shaded band shown in the left window of Figure \ref{pres}. Its upper and lower limit is obtained by using
the parameter set of $L=80$ MeV, $K_{\rm{sym}}=100$ MeV, $J_{\rm{sym}}=800$ MeV and $J_0=400$ MeV and the set of $L=40$ MeV, $K_{\rm{sym}}=-400$ MeV, $J_{\rm{sym}}=134$ MeV and $J_0=400$ MeV, respectively.
The individual roles of these parameters are examined by varying them independently in the right window. As one expects, the variation of $J_0$ is most effective in modifying the pressure at supra-saturation densities.

\section{Constraining the EOS of dense neutron-rich matter with observed properties of neutron stars}
\label{sec3}
With the EOSs prepared in the way described above, the mass (M)-radius (R) relationship of neutron stars is obtained by solving the Tolman-Oppenheimer-Volkov (TOV)
equations \citep{Tolman34,Oppenheimer39}
\begin{equation}\label{TOVp}
\frac{dP}{dr}=-\frac{G(m(r)+4\pi r^3P/c^2)(\epsilon+P/c^2)}{r(r-2Gm(r)/c^2)},
\end{equation}
\begin{equation}\label{TOVm}
\frac{dm(r)}{dr}=4\pi\epsilon r^2
\end{equation}
where $G$ is the gravitation constant, $c$ is the light speed and $m(r)$ is the gravitational mass enclosed within a radius $r$.
The dimensionless tidal deformability $\Lambda$ is related to the Love number $k_2$, neutron star mass M and radius R via
\begin{equation}
\Lambda = \frac{2}{3}k_2\cdot (R/M)^5.
\end{equation}
The $k_2$ is determined by the EOS thorough a differential equation coupled to the TOV equation \citep{Hinderer08,Hinderer10}.
More details about the formalism and code used in this work to calculate the $k_2$ can be found in, e.g., \citet{Fattoyev13,Fattoyev14}.

Within the 3-D parameter space of $J_0$, $K_{\rm{sym}}$ and $J_{\rm{sym}}$ and the 2-D parameter space of $L$ and $K_{\rm{sym}}$ under the conditions discussed in the previous section, using the observational data of $M_{\rm{max}}=2.01$ M$_\odot$, $10.62\leq R_{\rm{1.4}}\leq 12.83$ km and $\Lambda_{1.4}\leq 800$, we study how/what high-density EOS parameters are constrained in the following two subsections, separately. 

\subsection{Observational constraints in the $J_0-K_{\rm{sym}}-J_{\rm{sym}}$ EOS parameter space}
\label{sec3.2}

\begin{figure}[ht]
\begin{center}
  \includegraphics[width=10cm]{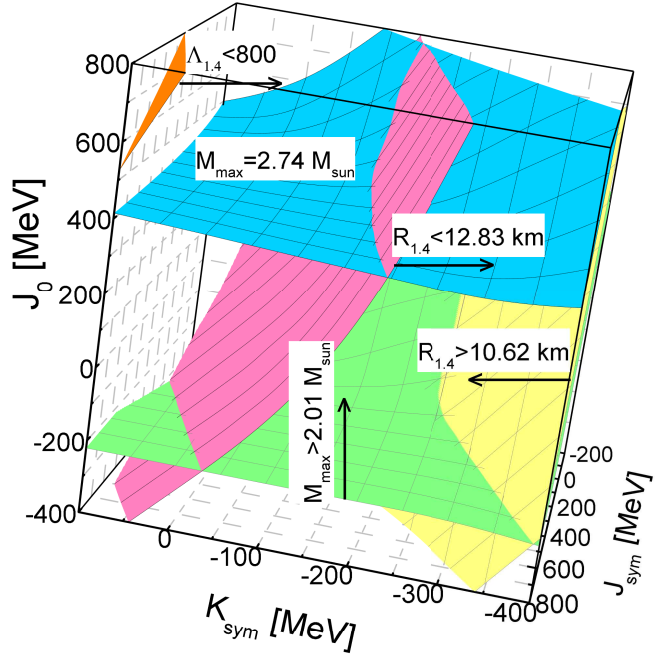}
  \caption{(color online) Observational constraints of the maximum mass of neutron stars and the radius of canonical neutron stars on the EOS of dense neutron-rich matter in the $K_{\rm{sym}}$, $J_{\rm{sym}}$ and $J_0$ parameter space. The green, magenta, yellow, blue and orange surfaces represent $M_{\rm{max}}=2.01$ M$_{\odot}$, $R_{1.4}=12.83$ km, $R_{1.4}=10.62$ km, $M=2.74$ M$_\odot$, and $\Lambda_{1.4}=800$, respectively.}\label{LTD}
\end{center}
\end{figure}
To be clear, we first emphasize again that in this case the following parameters are fixed at their currently known most probable values based on previous systematic surveys as discussed earlier: $K_0=240$ MeV, $E_{\rm sym}(\rho_0)=31.7$ MeV and $L=58.9$ MeV. We explore constraints on the EOS in the 3-D $K_{\rm{sym}}-J_{\rm{sym}}-J_0$ parameter space with the crust-core transition density consistently determined and the condition that $P_t\geq 0$. Technically, in exploring the 3-D parameter space in three loops we change the $J_0$ to reproduce a specific observational data at given values of $K_{\rm{sym}}$ and $J_{\rm{sym}}$ which are varied independently. 
Shown in Figure \ref{LTD} are the constant surfaces of neutron stars' maximum mass of $M_{\rm{max}}=2.01$ M$_{\odot}$ (green), radius of $R_{1.4}=12.83$ km (magenta) and $10.62$ km (yellow) as well as the upper limit of the dimensionless tidal deformability $\Lambda_{1.4}=800$ (orange) of canonical neutron stars, respectively. For clarity, a causality surface limiting M$\leq 2.4$ M$_{\odot}$ is not shown.

For ease of the following discussions, it is worth recalling first that, as shown in Eq. (\ref{PAEb}), the EOS of SNM $E_0(\rho)$, the symmetry energy $E_{\rm{sym}}(\rho)$ and the isospin asymmetry $\delta(\rho)$ at $\beta$ equilibrium are the three quantities together determining the total pressure in neutron stars. More specifically, the total pressure is proportional to $dE_0(\rho)/d\rho+\delta^2\cdot dE_{\rm{sym}}(\rho)/d\rho$.
While the $dE_0(\rho)/d\rho$ term is controlled by the $J_0$ parameter with a fixed $K_0$, the $dE_{\rm{sym}}(\rho)/d\rho$ and $\delta(\rho)$ are determined by the $K_{\rm{sym}}$ and $J_{\rm{sym}}$ parameters when the $L$ is fixed. Moreover, the symmetry energy contribution to the total pressure is weighted by $\delta^2(\rho)$. When the $E_{\rm{sym}}(\rho)$ is softer with smaller or negative $K_{\rm{sym}}$ and $J_{\rm{sym}}$ values, the system is more neutron-rich as shown in Figure \ref{Ksymeffect}. In particular, for extremely small $K_{\rm{sym}}$ and $J_{\rm{sym}}$ values (e.g., $K_{\rm{sym}}=-400$ MeV and $J_{\rm{sym}}=-200$ MeV), the $E_{\rm{sym}}(\rho)$ becomes negative and the $\delta(\rho)$ reaches its maximum of 1 at high densities. Then, the necessary contribution to the pressure from the $E_0(\rho)$ term will require a large $J_0$ value to support massive neutron stars.  At the other extreme, however, when both the $K_{\rm{sym}}$ and $J_{\rm{sym}}$ are strongly positive (e.g., $K_{\rm{sym}}=100$ MeV and $J_{\rm{sym}}=800$ MeV at the bottom left corner), the symmetry energy $E_{\rm{sym}}(\rho)$ is super-stiff and the $\delta(\rho)$ is very small as shown in Figure \ref{Ksymeffect}. The required $J_0$ to support massive neutron stars is then very small.

It is interesting to note several major features in this rather information-rich 3-D plot summarizing very extensive calculations. Let us first focus on the constant surface of $M_{\rm{max}}=2.01$ M$_{\odot}$. From the top right to the bottom left corner, the required $J_0$ first decreases quickly and then stays almost a constant with the increasing values of $K_{\rm{sym}}$ and $J_{\rm{sym}}$ from negative to positive. This feature is completely understandable based on the discussions in the previous paragraph. Namely, near the upper right corner, the symmetry energy is super-soft and the resulting $\delta(\rho)$ is close to 1. The weight  $\delta^2(\rho)$ of the symmetry energy contribution to the pressure is significant while the $dE_{\rm{sym}}(\rho)/d\rho$ value is small and may even be largely negative. To support neutron stars with the maximum mass of $M=2.01$ M$_\odot$, the value of $J_0$ has to be highly positive to compensate the small or overcome the possibly negative contribution from the symmetry energy. However, near the bottom left corner where the symmetry energy is super-stiff, the resulting $\delta^2$ at $\beta$ equilibrium becomes so small such that the symmetry energy contribution to the pressure is strongly suppressed. The necessary value of $J_0$ is therefore small and the constant surface of $M_{\rm{max}}=2.01$ M$_{\odot}$ becomes rather flat at small $J_0$ values. More quantitatively, the required minimum value of $J_0$ is about $-243$ MeV at $K_{\rm{sym}}=-400$  MeV and $J_{\rm{sym}}=800$ MeV.

For a comparison and see more clearly the role of $J_0$ parameter in determining the masses of neutron stars, a constant surface of $M_{\rm{max}}=2.74$ M$_{\odot}$ (sky blue) corresponding to the composite mass of GW170817 is also shown. It is seen that the overall shapes of the constant surfaces of $M_{\rm{max}}=2.01$ M$_{\odot}$ and 2.74 M$_{\odot}$ are rather similar. To support such a hypothetical neutron star would require a $J_0$ value above 400 MeV beyond its current uncertain range and the causality limit. While the fate of GW170817 is not completely determined yet, several analyses using observations of GW170817 combined with theories/simulations 
and the causality limit under some caveats have placed the upper bounds on neutron star masses in the range of $(2.16-2.28)$M$_\odot$ \citep{Margalit17,Shibata17, Rezzolla18, Ruiz18}, see, e.g., ref. \citep{Radice18} for a very recent review. For instance, \citet{Margalit17} used electromagnetic constraints on the remnant imposed by the
kilonova observations after the merger and the gravitational wave information predicted a maximum mass of $M_{\rm{max}}\leq 2.17$M$_\odot$ with 90\% confidence.
To support neutron stars with such masses,  the $J_0$ just needs to be slightly positive well within its current uncertain range. Thus, once the maximum mass is pinned down, it will put a stringent upper limit on the $J_0$ parameter.

In addition, it is also known that quadrupole deformations of neutron stars depend on the symmetry energy \citep{Plamen1,Plamen2,Fattoyev13,Fattoyev14,Fattoyev17,Plamen3}. Interestingly, \citet{Abbott17} inferred that the dimensionless tidal deformability of GW170817 has an upper limit of $\Lambda_{1.4}\leq800$ at the 90\% confidence level for the low-spin prior. However, it is seen in Figure \ref{LTD} that the constant surface of $\Lambda_{1.4}=800$ (orange) locates far outside the constant surface of $R_{1.4}=12.83$ km. Thus,  limits on the high-density EOS parameters from the $\Lambda_{1.4}\leq800$ constraint alone are presently much looser than the radius constraint extracted from analyzing the X-ray data. Nevertheless, the expected detection of gravitational waves from a large number of neutron star mergers has the potential to improve the situation.
\begin{figure}[ht]
\begin{center}
  \includegraphics[width=10cm]{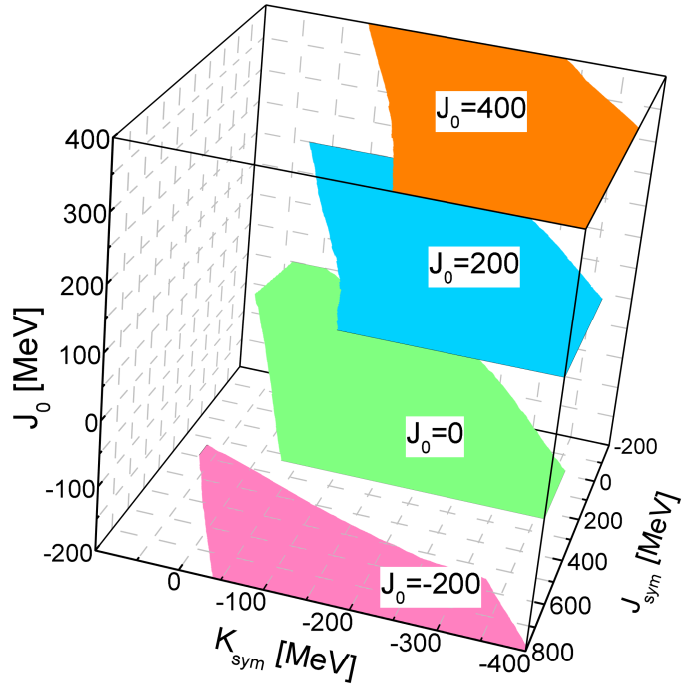}
  \caption{(color online) The allowed regions in the $K_{\rm{sym}}- J_{\rm{sym}}$ planes with $J_0=400$, $200$, $0$, and $-200$ MeV, respectively. }\label{constraint3D}
\end{center}
\end{figure}

Now, let us move to the EOS constraints from the radii of neutron stars. The radius of canonical neutron stars is known to depend strongly (weakly) on the nuclear symmetry energy (EOS of SNM) \citep{LiSteiner}. It is thus not surprising that the two constant surfaces of radius at $R_{1.4}=10.62$ km and $R_{1.4}=12.83$ km for canonical neutron stars are essentially vertical in Figure \ref{LTD}, indicating a weak dependence on the $J_0$ as one expects.
Indeed, they have significant dependences on both the $K_{\rm{sym}}$ and $J_{\rm{sym}}$ as indicated by the separation between the two constant-radius surfaces.
More quantitatively, the required values of $J_0$ in the constant surfaces of $R_{1.4}=10.62$ km and $R_{1.4}=12.83$ km decrease continuously with increasing $K_{\rm{sym}}$ and $J_{\rm{sym}}$. The two surfaces can be approximately described by $J_0=-585.64~ ({\rm{MeV}})-2.86K_{\rm{sym}}-1.00J_{\rm{sym}}$ and $J_0=182.54~ ({\rm{MeV}})-3.19K_{\rm{sym}}-0.60J_{\rm{sym}}$, respectively. With a fixed value of $L$, the nuclear pressure becomes stronger with increasing values of $K_{\rm{sym}}$ and $J_{\rm{sym}}$. Thus, the constant surface of $R_{1.4}=12.83$ km is on the left (having stiffer symmetry energies) of the $R_{1.4}=10.62$ km surface.

\begin{figure}[ht]
\begin{center}
  \includegraphics[width=10cm]{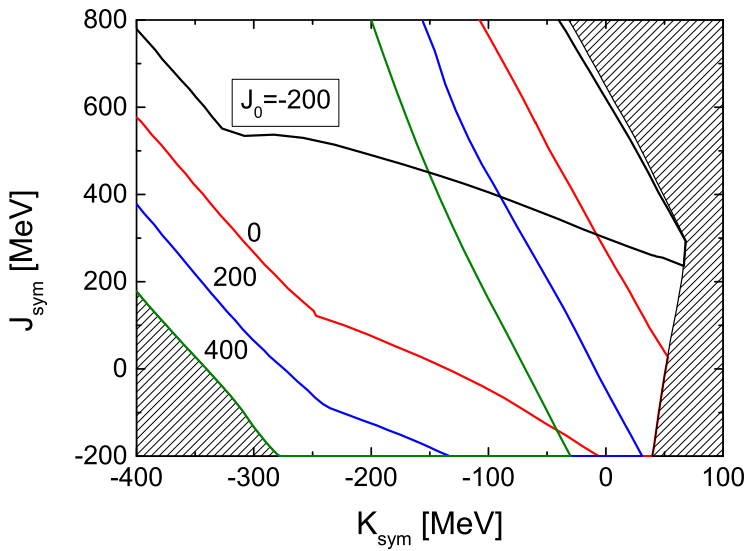}
  \caption{(color online) The constraining boundaries in the $K_{\rm{sym}}- J_{\rm{sym}}$ plane for a given $J_0$ value of $-200$, 0, 200, and $400$ MeV. The region insides the lines are allowed for a given $J_0$. The shadowed regions are excluded values of $K_{\rm{sym}}$ and $J_{\rm{sym}}$ for all $J_0$ values (see text). }\label{LT2D}
\end{center}
\end{figure}
As indicated by the arrows in Figure \ref{LTD}, the space enclosed by the three constant surfaces of $M_{\rm{max}}\geq 2.01$ M$_\odot$ and $10.62\leq R_{\rm{1.4}}\leq 12.83$ km are the EOS parameter space allowed by the astrophysical observations of the maximum mass and radii of neutron stars. The space is constrained from the bottom by the maximum mass, its intersections with the two limits on the radii set the boundary lines restricting the $K_{\rm{sym}}$ and $J_{\rm{sym}}$ at small $J_0$ values around $-200$ MeV. At higher $J_0$ values, the EOS parameter space is mainly bounded by the two radius constraints. To show the accepted EOS parameters more clearly, the allowed regions in the $K_{\rm{sym}}$ versus $J_{\rm{sym}}$ planes with $J_0=400$, $200$, $0$, and $-200$ MeV, respectively, are shown in Figure \ref{constraint3D}. The boundaries of these allowed regions in the $K_{\rm{sym}}- J_{\rm{sym}}$ planes are shown in Figure \ref{LT2D}. The regions insides the lines are allowed for a given $J_0$ specified. The shadowed regions are excluded values of $K_{\rm{sym}}$ and $J_{\rm{sym}}$ for all $J_0$ values. More specifically, the boundary of the right shadowed region can be divided into two parts based on the slopes. The upper one with a negative slope is obtained by requiring the crust-core transition pressure to always stay positive, i.e,  $P_t\geq 0$. It can be described approximately by the expression $J_{\rm{sym}}=-11.00K_{\rm{sym}}+457.71$ MeV. The lower one with a positive slope is obtained by the intersection line between the surfaces of $M_{\rm{max}}=2.01$ M$_\odot$ and $R_{\rm{1.4}}=12.83$ km. It can be fitted by the expression $J_{\rm{sym}}=7.68K_{\rm{sym}}-504.90$ MeV. This boundary sets an upper limit for $K_{\rm{sym}}$ at about $68$ MeV. The left shadowed region is excluded by the intersection line of $R_{\rm{1.4}}=10.62$ km and $J_0=400$ MeV in Figure \ref{LTD}. It can be fitted by the expression $J_{\rm{sym}}=-3.07K_{\rm{sym}}-1054.89$ MeV. 

Overall, within the framework of our analyses, using the maximum mass of neutron stars as well as the upper and lower limits of the radii of canonical neutron stars, the three EOS parameters are only limited in a space shown in Figure \ref{LTD} not completely closed with its boundaries partially given in Figure \ref{LT2D}. Obviously, data of more independent observables from either or/both astrophysics and nuclear physics are needed to determine the individual values of $K_{\rm{sym}}$, $J_{\rm{sym}}$ and $J_0$.

\subsection{Observational constraints in the $L-K_{\rm{sym}}$ EOS parameter plane}
\begin{figure}[ht]
\begin{center}
  \includegraphics[width=10cm]{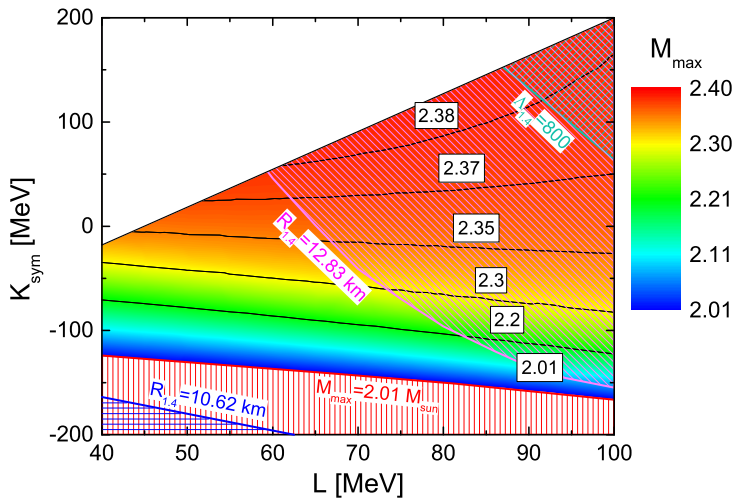}
  \caption{(color online)
Observational constraints of the maximum mass and radius of neutron stars on the EOS in the $L-K_{\rm{sym}}$ plane. The red, magenta, blue, and cerulean shadows represent regions where $M_{\rm{max}}\leq2.01$ M$_{\odot}$, $R_{1.4}\geq12.83$ km, $R_{1.4}\leq10.62$ km, and $\Lambda_{1.4}\leq800$, respectively. In the excluded white region the crust-core transition pressure $P_{t}\leq 0$.}\label{Mmax2D}
\end{center}
\end{figure}
Within the currently known uncertainty ranges of $J_0$ and $J_{\rm{sym}}$, one can parameterize the EOS with less parameters by setting $J_0=J_{\rm{sym}}=0$ in Eqs. (\ref{E0para}) and (\ref{Esympara}) as often done in the literature. Then, the two most poorly known parameters are the  $K_{\rm{sym}}$ and $L$. The latter is known to be around  $L\approx 58.7\pm 28.1 $ MeV as mentioned earlier. In this 2-D model framework, here we explore how/what the same three astrophysical constraints may teach us about the EOS of dense neutron-rich matter.

Shown in Figure \ref{Mmax2D} are contours of constant maximum masses of neutron stars in the $L-K_{\rm{sym}}$ plane. The red, magenta, blue, and  cerulean shadows represent regions where $M_{\rm{max}}\leq 2.01$ M$_{\odot}$, $R_{1.4}\geq 12.83$, $R_{1.4}\leq 10.62$ km, and $\Lambda_{1.4}\geq 800$, respectively. In the white excluded region, the crust-core transition pressure $P_{t}\leq 0$. It is seen that the maximum mass of neutron stars increases with increasing $K_{\rm{sym}}$ as one expects. However, it is rather insensitive to the variation of $L$ in the region considered. Again and obviously, the tidal polarizability $\Lambda_{1.4}\leq 800$ is much less restrictive than the radius constraint of $R_{1.4}\leq 12.83$ km. It is also seen that the $R_{1.4}\geq 10.62$ km constraint is covered by the constraint $M_{\rm{max}}\geq 2.01$ M$_{\odot}$. Consequently, the area bounded by the curves of $M_{\rm{max}}\geq 2.01$ M$_{\odot}$ and $R_{1.4}\leq 12.83$ km is the region allowed by the astrophysical observations considered. In this region, the maximum value of $K_{\rm{sym}}$ is about 52 MeV consistent with that found in the 3-D analyses in the previous subsection. Also in this region, the upper limit of the maximum mass is about 2.37 M$_{\odot}$ reached at $L\approx 60$ MeV. Probably incidentally, the latter is also currently the most probable value of $L$ based on the surveys of 53 analyses of existing data \citep{Li13,Oertel17}. Interestingly, the observed upper limit of the maximum mass is in good agreement with the findings by \citet{Fryer15}, \citet{Lawrence15} and \citet{Alsing17}. They found that the upper limit of the maximum mass of neutron stars is between $2.0$ and $2.5$ M$_\odot$. However, it is necessary to caution here that we have taken $J_0=J_{\rm{sym}}=0$ in the 2-D study. As we have discussed in the previous subsection, the value of $J_0$ affects significantly the maximum mass of neutron stars. Thus, without more precise knowledge about the $J_0$ parameter, the absolute maximum mass of neutron stars can not be pinned down. Based on our 2-D model analyses here, neutron stars more massive than 2.37 M$_{\odot}$ would require a positive value of $J_0$.

\section{Summary and outlook}
\label{sec4}
In summary, within both the 2-D and 3-D EOS parameter spaces limited by the existing constraints from terrestrial nuclear experiments, we studied how the astrophysical observations of
$M_{\rm{max}}>2.01\pm0.04$ M$_\odot$, $10.62<R_{\rm{1.4}}< 12.83$ km and $\Lambda_{1.4}\leq800$ all together constrain the EOS parameters of dense neutron-rich nucleonic matter.
We also investigated effects of the curvature $K_{\rm{sym}}$ of nuclear symmetry energy on the crust-core transition in neutron stars. The consistently calculated transition density in the $L-K_{\rm{sym}}$ plane
is used in constructing the EOS of neutron star matter from the surface to the core. The $K_{\rm{sym}}$ is found to affect significantly the crust-core transition density and pressure.
Fixing the $K_0$, $E_{\rm{sym}}(\rho_0)$ and $L$ at their most probably values determined mainly by terrestrial nuclear experiments, we explored the intersections of constant surfaces with $M_{\rm{max}}=2.01$ $\rm{M}_\odot$, $R_{\rm{1.4}}=10.62$ km, $R_{\rm{1.4}}=12.83$ km, and $\Lambda_{1.4}=800$, respectively, in the 3-D parameter space of $K_{\rm{sym}}$, $J_{\rm{sym}}$ and $J_0$.
The 3-D parameter space narrowed down significantly by the observational constraints is clearly identified. This helps guide nuclear theories for dense neutron-rich matter and related studies in terrestrial experiments.
However, to pin down the individual values of $K_{\rm{sym}}$, $J_{\rm{sym}}$ and $J_0$, data of additional independent observables from either astrophysical observations and/or laboratory experiments are needed. In particular, the skewness parameter $J_0$ of SNM largely controls the maximum mass of neutron stars. The 2-D EOS with $J_0=0$ is found sufficiently stiff to support neutron stars as massive as 2.37 M$_{\odot}$, while to support a hypothetical neutron star as massive as 2.74 M$_{\odot}$ (composite mass of GW170817) would require $J_0$ to be larger than about 400 MeV beyond its known limit. In both the 2-D and 3-D model frameworks considered in this work, the upper limit of the tidal deformability $\Lambda_{1.4}=800$ from the recent observation of GW170817 is found to provide upper limits on some EOS parameters consistent with but less restrictive than the existing constraints. In particular, its constraints on the symmetry energy parameters are far less restrictive than the observation of $10.62<R_{\rm{1.4}}< 12.83$ km from analyzing the X-ray data.

While the analyses and results presented here are useful in their own rights, some aspects of our present work should be improved in future studies.
In particular, more data and/or a better approach are necessary to determine precisely individual values of the high-density EOS parameters with quantified uncertainties. In the era of gravitational wave astronomy accompanied by the planned new experiments using advanced radioactive beam facilities around the world, we are very hopeful that more data will come soon. Moreover, even with the limited data available, more quantitative information about the high-density EOS parameters may be obtained by using a more robust statistical approach.  For example, we fixed the parameters describing the EOS near the saturation density at their most probable values mostly from terrestrial experiments. By doing so we eliminated any possible correlation between these (fixed) parameters and the high-density parameters $K_{\rm{sym}}$, $J_{\rm{sym}}$ and $J_0$ that we want to determine from astrophysical observations. A Bayesian analysis where the prior distribution of parameters encodes what is already known and the new observational data refines such prior knowledge is particularly well suited for inferring the posterior probability density distributions of all the EOS parameters. Such a study using the masses and radii of 14 neutron stars extracted from Chandra X-Ray observations, the maximum masses of neutron stars from GW170817 and earlier observations of neutron stars as well as all the information from terrestrial experiments about the low-order EOS parameters will be carried out and reported elsewhere.

\acknowledgments
We thank F. J. Fattoyev for providing us the code to calculate the dimensionless tidal deformability. We would also like to thank Alessandra Corsi, Benjamin Owen and Renxin Xu for very helpful discussions as well as 
Ang Li and Bing Zhang for useful communications. NBZ is supported in part by the China Scholarship Council. BAL acknowledges the U.S. Department of Energy, Office of Science, under Award Number DE-SC0013702, the CUSTIPEN (China-U.S. Theory Institute for Physics with Exotic Nuclei) under the US Department of Energy Grant No. DE-SC0009971 and the National Natural Science Foundation of China under Grant No. 11320101004. JX is supported in part by the Major State Basic Research Development Program (973 Program) of China under Contract Nos. 2015CB856904 and 2014CB845401, the National Natural Science Foundation of China under Grant Nos. 11475243 and 11421505, the ``100-talent plan'' of Shanghai Institute of Applied Physics under Grant Nos. Y290061011 and Y526011011 from the Chinese Academy of Sciences, and the Shanghai Key Laboratory of Particle Physics and Cosmology under Grant No. 15DZ2272100.

\end{document}